\documentclass[%
 reprint,
 amsmath,amssymb,
 aps,
]{revtex4-2}

\usepackage{graphicx}
\usepackage{dcolumn}
\usepackage{bm}

\usepackage{amssymb, amsmath, amsthm}
\usepackage{xcolor}
\usepackage{graphicx}
\usepackage[protrusion=true,expansion=true]{microtype}
\usepackage{times}
\usepackage{hyperref}
\hypersetup{
    pdftitle={main},    
    pdfauthor={},
    colorlinks=true,
    citecolor=blue,
    linkcolor=blue,
    urlcolor=blue
  }
\usepackage{comment}
\usepackage{blkarray}
\usepackage{mathtools}
\usepackage{tikz}
\usetikzlibrary{decorations.pathreplacing}
\usepackage{dsfont}

\usepackage{graphicx}
\usepackage{subfigure}
\usepackage{dcolumn}
\usepackage{bm}

\usepackage[makeroom]{cancel}
\usepackage{latexsym}
\usepackage{longtable}
\usepackage{epsfig}
\usepackage{graphicx,bbm,psfrag}
\usepackage{amssymb,amsmath,amsthm,booktabs,mathtools}
\usepackage{bm}
\usepackage{color}
\usepackage{dutchcal}
\usepackage{hyperref}

\hypersetup{
    colorlinks,
    citecolor=red,
    linkcolor=blue,
}

\newcommand{\bc}{\begin{center}}
\newcommand{\ec}{\end{center}}
\def\ba#1{\begin{array}{#1}\displaystyle}
\newcommand{\ea}{\end{array}}

\newcommand{\beq}{\begin{equation}}
\newcommand{\eeq}{\end{equation}}
\newcommand{\beqa}{\begin{eqnarray}}
\newcommand{\eeqa}{\end{eqnarray}}

\newcommand{\bi}{\begin{itemize}}
\newcommand{\ei}{\end{itemize}}

\renewcommand{\v}[1]{\boldsymbol{#1}}

\def\b#1{\bar{#1}}
\def\frc#1#2{\frac{#1}{#2}}

\newcommand{\p}{\partial}

\newcommand{\bra}{\langle}
\newcommand{\ket}{\rangle}

\newcommand{\R}{{\mathbb{R}}}

\newcommand{\Tr}{{\rm Tr}}

\newcommand{\ep}{\epsilon}

\newcommand{\ri}{{\rm i}}

\newcommand{\dd}{{\rm d}}

\DeclareMathOperator{\sgn}{sgn}

\newcommand{\titleinfo}{
Full counting statistics after quantum quenches as hydrodynamic fluctuations
}
\begin{document}

\preprint{APS/123-QED}

\title{\titleinfo
}

\author{Dávid X. Horváth}
\affiliation{Department of Mathematics, King’s College London, Strand WC2R 2LS, London, U.K.}

\author{Benjamin Doyon}
\affiliation{Department of Mathematics, King’s College London, Strand WC2R 2LS, London, U.K.}

\author{Paola Ruggiero}
\affiliation{Department of Mathematics, King’s College London, Strand WC2R 2LS, London, U.K.}

\date{\today}

\begin{abstract}
The statistics of fluctuations on large regions of space encodes universal properties of many-body systems. At equilibrium, it is described by thermodynamics. However, away from equilibrium such as after quantum quenches, the fundamental principles are more nebulous. In particular, although exact results have been conjectured in integrable models, a correct understanding of the physics is largely missing.
In this paper, we explain these principles, taking the example of the number of particles lying on a large interval in one-dimensional interacting systems. These are based on simple hydrodynamic arguments from the theory of ballistically transported fluctuations, and in particular the Euler-scale transport of long-range correlations. Using these principles, we obtain the full counting statistics in terms of thermodynamic and hydrodynamic quantities, whose validity depends on the structure of hydrodynamic modes and on the fluctuations in the initial state. In fermionic-statistics interacting integrable models with a continuum of hydrodynamic modes, such as the Lieb-Liniger model for cold atomic gases, the formula reproduces previous conjectures, but is in fact not exact: it gives the correct cumulants up to, including, order 5, while long-range correlations modify higher cumulants. In integrable and non-integrable models with two or less hydrodynamic modes, the formula is expected to give all cumulants.  
\end{abstract}

\maketitle


\section{Introduction} 
The problem of understanding fluctuations in many-body systems has received a large amount of attention recently. {\em At equilibrium}, fluctuations on large regions are fully determined by thermodynamic functions. For instance, in an infinitely large system of particles at finite density, the cumulants $C_n=\bra Q^n\ket^{\rm c}$ of an extensive charge $Q$ (say the number of particles) supported on a region of space of finite but large volume $V$ are generated by the difference of Landau free energies $f(\b\mu)$ as $F(\lambda) = \sum_{n=1}^\infty \lambda^n C_n/n! = V(f(\b\mu)-f(\b\mu+\lambda))$ (here only the dependence on $\b\mu $, which is the chemical potential divided by the temperature, associated to the charge $Q$ is shown). This is often referred to as the {\em full counting statistics} (FCS); its Legendre-Fenchel transform gives the probability distribution, peaked around the average $\bra Q\ket = \b {\mathcal Q}$ and given by the relative entropy (Kullback–Leibler divergence $D_{\rm KL}$): $P(Q = \mathcal Q) \asymp e^{-V D_{\rm KL}(\rho_{\mathcal Q}||\rho_{\b {\mathcal Q}}))}$, where $\rho_{\mathcal Q}$ is the grand-canonical distribution biased to the average $\bra Q\ket_{\rho_{\mathcal Q}} =\mathcal Q$.
This applies both to classical and quantum systems, and holds in any non-equilibrium steady states (NESS) with short-range correlations.

Dynamical quantities, such as the total amount of charge going through an interface in a finite but large time $T$, go beyond thermodynamics. But in some cases, it is known that the large-deviation form, as above, of their fluctuations only requires the knowledge of the hydrodynamic theory associated to the many-body systems. If the charge admits ballistic transport, one uses Euler hydrodynamics via ballistic fluctuation theory (BFT) \cite{BFT1-Doyon_2018, BFT2-Doyon_2019, BFT3-Myers_2020, BFT4-Fava_2021, BFT5-Perfetto_2021},  or the more general ballistic macroscopic fluctuation theory (BMFT) \cite{doyon2022ballistic}; while if transport is diffusive, macroscopic fluctuation theory (MFT) \cite{MFT1, MFT2} relates fluctuations to the diffusivity parameters. In non-stationary but {\em slowly-varying} states, where local entropy maximization has occurred, hydrodynamic principles still hold, and BMFT or MFT apply \cite{Doyon_2020}. It is worth noting that in certain situations, fluctuations are anomalous and the large-deviation principle is broken \cite{NoLD1-Krajnik_2022, NoLD2-Krajnik_2022, DeNardisSqueezed2023, NoLD3-Krajnik_2024, Gopalakrishnan_2024, mcculloch2024ballisticmodessourceanomalous}.

Truly {\em far from equilibrium}, however, much less is known. A common protocol is that of quenches: sudden changes of coupling constants or other parameters determining the dynamics. It is known that entropy maximization constrained by the extensive charges -- (generalized) thermalization -- occurs at long times \cite{kinoshita2006quantum, langen2015experimental, GGE1-Polkovnikov_2011, Calabrese_2011_CEFPRL, Calabrese_2012_CEF1,Calabrese_2012_CEF2, GGE2-calabrese2016introduction, GGE3-Vidmar_2016}.
But what about fluctuations? For instance, how do conserved charges on a volume $V$ fluctuate after a time $T$? This is non-trivial: for any $T$ that is not infinitely large compared to the diameter of the region, the long-time steady state {\em does not} describe large-scale fluctuations, because it misses long-range correlations that are known to emerge \cite{Doyon_LongRangeCorrelations}. Further, the fluctuations are dynamical, but common hydrodynamic fluctuation arguments such as linear response are not applicable, because quenches are far from equilibrium.

In this paper,
we propose a {\em simple non-linear hydrodynamic argument} for large-scale fluctuations after quenches, in models with ballistic transport (such as gases, with a conserved momentum). This is based on the physical principle of hydrodynamic transport via BFT. The main idea is that one must account for {\em the transport of long-range correlations produced by the quench}. We concentrate on the fluctuation of the total number of particles in quantum gases on a large interval, but the argument is relatively general. We express the FCS in a physically transparent way, solely in terms of thermodynamic and hydrodynamic quantities constructed from long-time steady states associated to the quench, and to a biased version of it. The formula is valid up to a certain order $n$ of the generated cumulants $C_n$, which may be infinite, and which depends on the structure of the hydrodynamic modes.
It can be applied to any interacting many-body system, as long as the corresponding thermodynamics, Euler hydrodynamics, and required steady states, are known. 
This is the case for ``integrable quenches," that is, quenches from integrable initial states (as defined in \cite{piroli2017what}) in integrable systems, our main focus -- see \cite{LongPaper} for the more general setup.
In integrable quenches of fermionic-statistics integrable systems with a continuum of hydrodynamic modes, such as the much-studied BEC quench of the paradigmatic Lieb-Liniger (LL) model describing cold atomic quantum gases \cite{denardis2014solution,De_Nardis_2014, Nardis_2015, De_Nardis_2016,Bastianello_2018,Perfetto_2019, FCSQGases}, the formula reproduces the cumulants up to, including, order 5; higher cumulants are affected by long-range correlations in a way that cannot yet be evaluated. In models, integrable or not, with only two hydrodynamic modes of opposite-sign velocities, or one hydrodynamic mode, the formula is expected to give all cumulants.

In integrable quenches, using generalized hydrodynamics (GHD) \cite{castroalvaredo2016emergent, bertini2016transport} and the Quench Action method \cite{caux2013time, caux2016quench} to evaluate exactly the required long-time steady states, we show that our general formula specializes to a formula recently conjectured using space-time swap techniques \cite{Bertini_2023_tFCS, Bertini_2024_tFCS}. This was verified in the quantum rule-54 cellular automaton \cite{Bertini_2023_tFCS, Bertini_2024_tFCS} (which admits two hydrodynamic velocities). Moreover, numerical data in the XXZ model \cite{Bertini_2024_tFCS} (with a continuum of velocities) show agreement at small values of $\lambda$, but errors appear at large values. Notwithstanding finite numerical effects, this is compatible with our result that the formula receives corrections at higher orders in $\lambda$.

The organization of the paper is as follows. In Sec. \ref{Sec:SetupAndMainResults} we introduce the setup and the non-equilibrium protocol under investigation, and we present the hydrodynamic picture for the initial behaviour of the FCS as well as its prediction. 
The core idea is that the early-time dynamics of charge FCS after a quench can be extracted from current fluctuations in a non-equilibrium steady state associated with a bipartite formulation of the original homogeneous quench problem. In many cases, this formulation is technically simpler to analyse.
In Sec. \ref{Sec:Derivation} we present an intuitive derivation for the aforementioned picture. We also highlight important assumptions, and comment on their plausibility. In Sec. \ref{TClusteringViolationMain} we then show that an important assumption on the decay of current autocorrelation functions may not hold in certain integrable systems. This can be due to what we call "hydrodynamic waft effect" \cite{LongPaper}, which transports initial state correlation along curved hydrodynamic trajectories building up stronger current-current correlations. Interestingly, this effect modifies the results of the simple hydrodynamic picture at large orders in the counting field.
These three self-contained sections are the core of the paper and contain the main ideas and the key findings. 

Secs. \ref{Sec:TBA} and \ref{Sec:Curved} target important, but more specific aspects of our main findings.
In Sec. \ref{Sec:TBA} we introduce some basic machinery of integrable models. As said, for these models and for certain quenches, it is possible to provide explicit predictions for the initial behaviour of the FCS. Additionally, such systems can be used to go beyond the proposed hydrodynamic picture and reveal sources of corrections.  In this section we shall discuss mostly how the flow equations for current fluctuations can be evaluated in the arising bipartite quench geometry for such models. For the sake of concreteness we introduce the Lieb-Liniger model, one of the simplest integrable models yet exhibiting experimental relevance, and a particular integrable quench, when the initial state is a Bose--Einsten condensate (BEC).

In Sec. \ref{Sec:Curved}, we first discuss the trajectories of hydrodynamic normal modes in the modified bipartite quench problem. Focusing on Euler-scale hydrodynamics and exploiting integrability, we show that, in generic integrable models, the trajectories associated with opposite quasi-particle momenta become curved at small but finite momenta.
This finding has notable consequences as discussed in
the remainder of the section.
Subsec. \ref{Sec:BMFT} develops a picture based on the BMFT \cite{doyon2022ballistic} 
in which the dynamics is described as the ballistic propagation of initial fluctuations.
In this subsection, we first infer the initial fluctuations of conserved densities in integrable initial states. This is achieved by studying such initial states in free systems, but given the specific structure of such integrable states, the initial large-scale correlations are expected to be similar in interacting cases. Using the form of these correlations, we then demonstrate that their ballistic propagation yields corrections to the naive predictions of the hydrodynamic picture, which, surprisingly, affect only high order cumulants.

Finally, our summary and conclusions are presented.

\section{Setup and main result} 
\label{Sec:SetupAndMainResults}
Consider a many-particle integrable quantum system in the thermodynamic limit on the line, with Hamiltonian $H$, and an integrable initial state $|\Psi\ket$. In quantum quenches, typically $|\Psi\ket$ is the ground state for a Hamiltonian $H'$ obtained from $H$ by changing an interaction strength. Let $Q$ be the total number of particles, and $Q|_0^X = \int_0^X \dd x\,q(x)$ that on the interval $[0,X]$. We are interested in the FCS
\beq
    F(\lambda) = \log \Big(\bra \Psi| e^{\lambda Q|_0^X(T)}|\Psi\ket\Big)
\eeq
where $Q|_0^X(T) = e^{\ri HT} Q|_0^X e^{-\ri HT}$. Its Taylor expansion in $\lambda$ gives the cumulants $C_n = \big\bra \Psi\big| \big(Q|_0^X(T)\big)^n\big|\Psi\big\ket^{\rm c}$ of $Q|_0^X(T)$ in the state $|\Psi\ket$ (here and below ${}^{\rm c}$ denotes the connected part). We are looking for the ballistic scaling regime $X,T\propto \ell\to\infty$ with $X/T = \alpha = {\rm const}$, and the scaled cumulants $c_n = C_n/\ell$ (which have a finite limit) \footnote{In the following, the quantities we discuss will be defined in the sequential limit $\lim_{X\to \infty}\lim_{T\to \infty}$ and $\lim_{T\to \infty}\lim_{X\to \infty}$. This is expected to be equivalent to the limit ${\alpha \to 0}$ and $\alpha \to \infty$, respectively, as we are in the ballistic regime.}. Taking $\alpha \to 0$ is simple: this is the limit $T\to\infty$, followed by the asymptotic $X\to\infty$. As $e^{\lambda Q|_0^X}$ is supported on a finite interval (that is, ``local"), we may use the long-time steady state of the quench $e^{-W_\Psi}$:
\beq\label{longtimepsi}
    \langle \Psi |O(t)|\Psi\rangle\rightarrow\Tr[O \,e^{-W_\Psi}]/\Tr[e^{-W_\Psi}] \quad (t\to\infty,\ O\mbox{ local}).
\eeq
Generally, such state is a generalized Gibbs Ensemble (GGE), i.e., $W_\Psi=\sum_i\beta^iQ_i$ for a basis of infinitely-many (quasi) local conserved charges and chemical potentials. 
Therefore the result is obtained by biasing the Landau free energy as described above, $f_{W_\Psi}(\b\mu) = -\log \Tr (e^{-W_\Psi + \b\mu Q})$:
\beq\label{Falpha0}
    F(\lambda)|_{\alpha=0} \sim X (f_{W_\Psi}(0) - f_{W_\Psi}(\lambda)) \quad (X\to\infty).
\eeq

The more interesting region is when  $X \gg T\gg 1$, i.e. that of ``small macroscopic time" or $\alpha\to\infty$, where the dynamics is non-trivial. This is difficult, because of the presence of {\em hydrodynamic long-range correlations} emanating from the quench \cite{PhysRevA.89.053608,del_Vecchio_del_Vecchio_2024}. Such correlations are often interpreted in a quasi-particle picture, where pairs of opposite-momentum, correlated excitations are emitted \cite{Calabrese_2005, Alba_2017}. In integrable quenches, only such pairs are emitted \cite{piroli2017what}, simplifying the structure of long-range correlations and allowing for the explicit characterization of the GGE \cite{caux2013time, caux2016quench}. We will restrict ourselves to these for ease of the discussion, but we note that in general, the concept of quasi-particle is replaced by that of hydrodynamic mode; with this understanding, we will also comment on what happens away from integrability.

\begin{figure}
    \centering
    \includegraphics[width=0.8\linewidth]{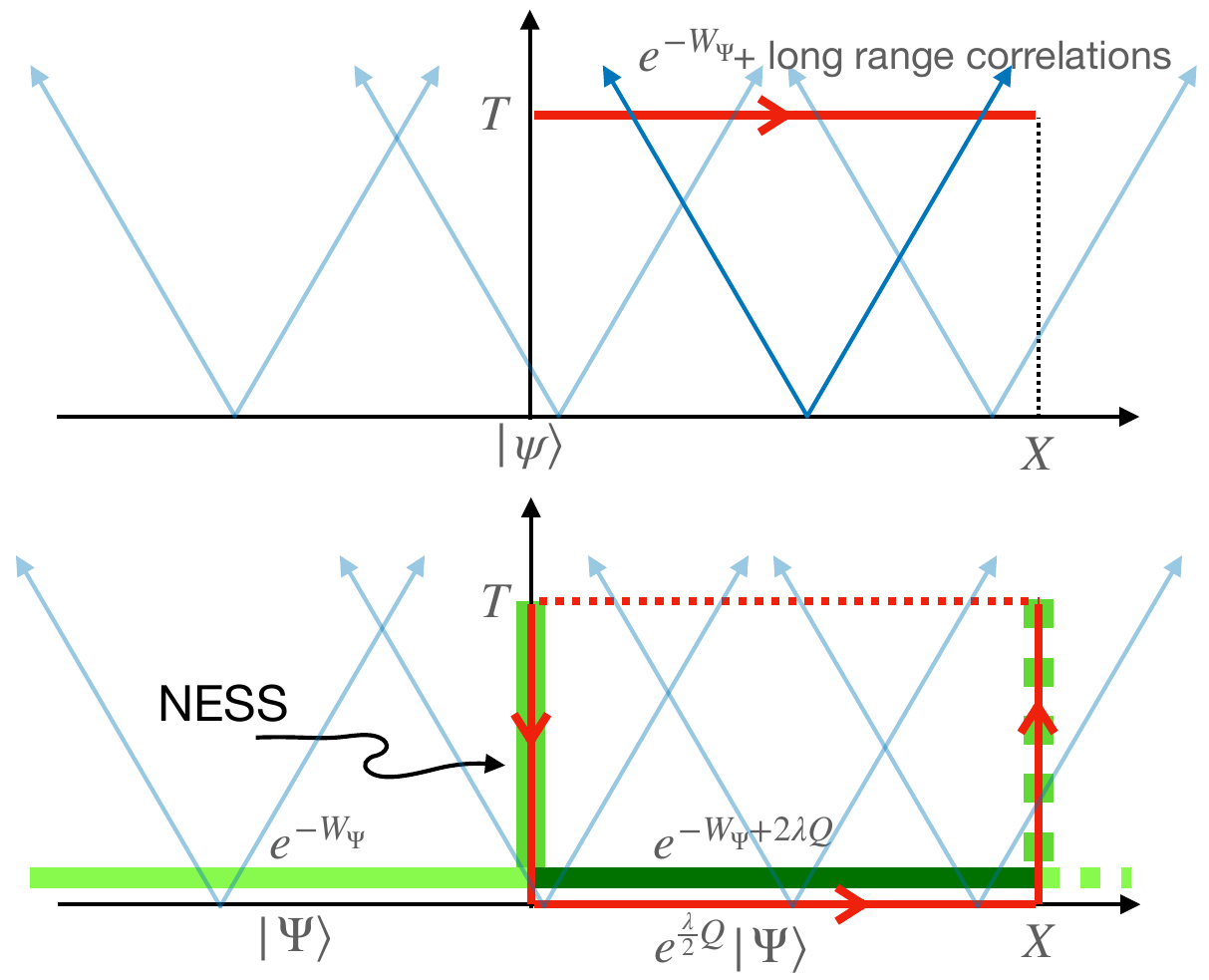}
    \caption{Illustration of the universal hydrodynamic principles for full counting statistics of the total charge $Q|_0^X(T)$ on length $X$ at time $T$ after a quench. The red contour represents this charge as an integrated density $\int_0^X \dd x\,q(x,T)$ in the generator $F=\log\langle\Psi| e^{\lambda Q|_0^X(T)}|\Psi\rangle$. The contour is deformed using the continuity equation in order to avoid long-range correlations emanating from the initial state and ``connecting" any parts of it (the darker blue arrows). Green bars represent steady states: from the biased quench on the right (locally $e^{\frc\lambda2 Q}|\Psi\ket$) and the original quench on the left (locally $|\Psi\ket$), and the non-equilibrium steady state (NESS) from the partitioning protocol they induce. The specific steady state $e^{-W_\Psi + 2\lambda Q}$ is our general result for integrable quenches. The partitioning protocols at $0$ and $X$ are independent for $X$ large enough. As every part of the deformed contour is uncorrelated, they give rise to independent contributions: the current FCS at $x=0, X$ (twice the same contribution, under parity symmetry) giving the linear growth of $F(\lambda)$ with $T$, and the charge FCS at $t=0$ giving the non-dynamical contribution.}
    \label{fig:quench}
\end{figure}

In order to illustrate the effects of hydrodynamic long-range correlations, consider the second cumulant $C_2 = \ell c_2$:
\begin{align}
    \ell c_2 & = \int_{0}^{X} \dd x\int_{0}^{X}\dd x'\,
    \bra\Psi|q(x,T)q(x',T)|\Psi\ket^{\rm c} \nonumber\\
    &=\ell^2\int_{0}^{\b X} \dd \b x\int_{0}^{\b X}\dd \b x'\,
    \bra\Psi|q(\ell \b x,\ell \b T)q(\ell \b x',\ell \b T)|\Psi\ket^{\rm c}\label{c2example}
\end{align}
where $X = \ell \b X,\,T = \ell\b T$. Hydrodynamic long-range correlations arise as a nonzero ``$E$-terms" \cite{Doyon_LongRangeCorrelations} in
\begin{equation}
    \ell \bra\Psi|q(\ell \b x,\ell \b T)q(\ell \b x',\ell \b T)|\Psi\ket^{\rm c}
    \sim c_2|_{W_\Psi}\delta(\b x-\b x')
    + E_\Psi(\b x,\b x',\b T)
\end{equation}
where $c_2|_{W_\Psi}$ is the scaled second cumulant in the GGE
$e^{-W_\Psi}$ (not to be confused with $c_2$). Clearly, then, the long-range correlations influence the result of $c_2$ in \eqref{c2example}, making it different from $c_2|_{W_\Psi}$ and time-dependent. See the illustration in Fig.~\ref{fig:quench} to understand where, na\"ively, particle-pair long-range correlations are located: with increasing $T$ the number of correlated pairs within the interval $[0,X]$ decreases linearly with $T$.

We now argue, using the continuity equation for the particle density $\p_t q + \p_x j=0$,
that the following formula for $F(\lambda)$ {\em gives the scaled cumulants $c_n$} in the limit $\alpha\to\infty$, up to a certain order $n$ (which may be infinite) depending on hydrodynamic properties. At $\alpha=\infty$ we first take the asymptotic $X\to\infty$, followed by the asymptotic $T\to\infty$ (corresponding to the regime $X \gg T\gg 1$), and as we shall shortly demonstrate, the FCS can be written as (for precise definitions see \cite{LongPaper})
\begin{align}\label{main}
    F(\lambda) \sim &\, X [f_{W_\Psi}(0)-f_{W_\Psi}(2\lambda)+o(X^0)]+ 2Tf_{\rm dyn}(\lambda) \\
    &\qquad\qquad\qquad\qquad(X\to\infty,\mbox{ then } T\to\infty).\nonumber
\end{align}
The corrections $o(X^0)$ (vanishing as $X\to\infty$) {\em are independent of $T$ for all $X$}. The dynamical free energy $f_{\rm dyn}$ in general depends on $\alpha$, and its limit $f_{\rm dyn}(\lambda) = \lim_{\alpha\to\infty} f_{\rm dyn}(\lambda,\alpha)$ is given by {\em the FCS for total current fluctuations} 
\begin{align}\label{fdyn}
    & f_{\rm dyn}(\lambda) =\\
    &\  \lim_{T\to\infty} \frc1T \log \Big(\Tr\, e^{-W_{\rm NESS}^\lambda} e^{\lambda \int_0^{T} \dd t\,j(0,t)}/
    \Tr\, e^{-W_{\rm NESS}^\lambda}\Big)\nonumber
\end{align}
in the state $e^{-W_{\rm NESS}^\lambda}$. This is the unique NESS for a $\lambda$-dependent {\em partitioning protocol}, where the initial state is $e^{-W_\Psi}$ on the left and $e^{-W_\Psi^\lambda}$, with $W_\Psi^\lambda = W_\Psi-2\lambda Q$, on the right, namely 
\begin{align}\label{partitioning}
    & \Tr [O(t)\,e^{-W_{\Psi}|_{-\infty}^0 - W_{\Psi}^\lambda|_0^\infty}]
    \to\Tr [O\,e^{-W_{\rm NESS}(\lambda)}]
    \quad (t\to\infty).
\end{align}
A result that we show in the App. \ref{QAComputation} is the local convergence to the $W_{\Psi}^{\lambda}$ GGE of the biased state on the right, i.e.,
\beq\label{longtimepsiQ}
    \frc{\langle \Psi |e^{\frc\lambda2 Q}O(t)e^{\frc\lambda2 Q}|\Psi\rangle}{\langle \Psi |e^{\lambda Q}|\Psi\rangle}\rightarrow \frc{\Tr[O \,e^{-W_{\Psi}^{\lambda}}]}{\Tr[e^{-W_{\Psi}^{\lambda}}]} \quad (t\to\infty).
\eeq
Note that the coefficient of $Q$ in $W_{\Psi}^{\lambda}$ is $2\lambda$. This ``doubling" phenomenon is specific to integrable quenches, and encodes the effects of correlated pairs \cite{Alba_2017QA,Bertini_2024_tFCS, FCSQGases}. 

Two remarks are in order. First, the non-dynamical part in Eq. \eqref{main}, proportional to $X$, is similar to the case $\alpha=0$, but with $2\lambda$ instead of $\lambda$. This was shown in \cite{Bertini_2024_tFCS, FCSQGases}, but it also follows from \eqref{longtimepsiQ} (see App. \ref{QAComputation}). We stress that the appearance and the precise form of $f_{W_\Psi}(2\lambda)$ is specific to integrable quenches.
Second, the dynamical part in Eq. \eqref{main}, given by Eq. \eqref{fdyn}, involves the FCS of the total time-integrated current in the NESS associated with a partitioning protocol between the two GGEs, where one side is biased by the charge with normalized chemical potential $2 \lambda$. The NESS may be evaluated by standard hydrodynamic arguments as done in \cite{castroalvaredo2016emergent,bertini2016transport} for integrable systems, and the FCS $f_{\rm dyn}(\lambda)$ also follows from hydrodynamics using the framework of BFT \cite{BFT1-Doyon_2018, BFT2-Doyon_2019, BFT3-Myers_2020}. For non-integrable systems, $W_\Psi^\lambda = W_\Psi - 2\lambda Q$ is replaced by some $W_\Psi^\lambda$ representing the long-time steady state for a quench from the biased state $e^{\frc\lambda2 Q}|\Psi\ket$ (as in Eq.~\eqref{longtimepsiQ}), and $f_{W_\Psi}(2\lambda)$ is replaced by $f_{W_\Psi^\lambda}$ in \eqref{main}; note that the BFT can still be applied. Hence, knowing the long-time quench steady states for $|\Psi\ket$ and its biased version $e^{\frc\lambda2 Q}|\Psi\ket$ (known in integrable systems), {\em every object in \eqref{main} is determined by hydrodynamics and thermodynamics}, thus readily calculable. Eqs.~\eqref{main}-\eqref{partitioning}, with this understanding, is our main result.
In Sec. \ref{Sec:TBA} we specify these equations explicitly for the example of the LL model subject quenching from a BEC state. Finally, we note that the fact that \eqref{main} splits into an $X$ and a $T$-dependent bit, and that the latter is defined by current fluctuations is a general statement not necessitating integrability assumptions (cf. \cite{LongPaper}).

\section{Derivation}
\label{Sec:Derivation}
The derivation uses basic properties of many-body systems, and is illustrated in Fig.~\ref{fig:quench}. From the continuity equation, $Q|_0^X(T) = \int_0^X \dd x\,q(x,T)
    =\int_0^T \dd t\,j(0,t)
    + \int_0^X \dd x\,q(x,0)
    - \int_0^T \dd t\,j(X,t)$,
and thus in a first step:
\begin{align}\label{secondstep}
    &\bra \Psi|e^{\lambda Q|_0^X(T)}|\Psi\ket = \Big(\bra\Psi|e^{\lambda Q|_0^{X/2}}|\Psi\ket\Big)^2 \times\\
    &\qquad \frc{\bra \Psi|e^{\frc\lambda2 Q|_0^{X/2}}
    e^{\lambda J_0|_0^T}
    e^{\frc\lambda2 Q|_0^{X/2}}
    e^{\frc\lambda2 Q|_{X/2}^{X}}
    e^{-\lambda J_X|_0^T}
    e^{\frc\lambda2 Q|_{X/2}^{X}}|\Psi\ket}{
    \Big(\bra\Psi|e^{\lambda Q|_0^{X/2}}|\Psi\ket\Big)^2
    }\nonumber
\end{align}
where $J_x|_0^T = \int_0^T \dd t\,j(x,t)$. In the second line, we have separated exponentials as $e^{A+B} \approx e^Ae^B$ by using the Baker-Campbell-Hausdorff (BCH) formula, neglecting all commutators of the operators $A_1=Q|_0^{X/2}$, $A_2=Q|_{X/2}^{X}$, $A_3=J_0|_0^T$ and $A_4=J_X|_0^T$. Neglecting these commutators means that the leading effects in the ballistic limit are purely classical. It is justified as follows. By locality $[A_1,A_2] = O(X^0)$ independently of time. By the Lieb-Robinson bound, or more generally, the sufficient decay of high-velocity contributions to operator evolutions, $A_3, A_4$ are (with exponential accuracy) supported on regions of lengths $vT$ (for some $v$), thus in the limit $X\to\infty$ we have $[A_1,A_4],\, [A_2,A_3],\, [A_3,A_4]\to0$. Concerning $[A_1,A_3]$ and $[A_2,A_4]$, we currently do not know of general results leading to sharp bounds. However, using the fact that $Q$ is ``ultra local", we propose (see App. \ref{App:Commutator}) general ideas based on thinness arguments \cite{Ampelogiannis2023} which support the fact that they can also be neglected, confirmed by explicit free-fermion calculations \cite{LongPaper}. The BCH formula in fact involves multiple-commutators, which can be analyzed in similar ways.

In a second step, we use results from integrable quenches obtained by correlation matrix \cite{Bertini_2024_tFCS} or quench action methods \cite{caux2013time, caux2016quench, FCSQGases}, giving for the first factor on the right-hand side of \eqref{secondstep}
\begin{equation}
\log \bra\Psi|e^{\lambda Q|_0^{X/2}}|\Psi\ket^2 
    \sim X  (f_{W_\Psi}(0) - f_{W_\Psi}(2\lambda)).
\end{equation}
This shows the term proportional to $X$ in \eqref{main}. This doubling can be understood in terms of the particle-pair picture: as the free energy does not account for long-range correlations, it only encodes the distribution of single members of each pair. But in the state $e^{\frc{\lambda}2 Q|_0^X}|\Psi\ket$, the modified quantum amplitude gives rise to a modified probability $e^{\lambda Q|_{\rm pair}}$ for each pair, hence, at later time when members are separated, for each single member of a pair, where a doubling occur: $Q|_{\rm pair} = 2Q|_{\rm single\,particle}$. See App. \ref{QAComputation}  for the explicit calculation as well as the work \cite{LongPaper} discussing general but clustering initial states.

In a third step, the second factor on the right-hand side of \eqref{secondstep} is factorized as
\begin{equation}
    \Bigg(\frc{\bra \Psi|e^{\frc\lambda2 Q|_0^{\infty}}
    e^{\lambda J_0|_0^T}
    e^{\frc\lambda2 Q|_0^{\infty}}
    |\Psi\ket}{
    \bra\Psi|e^{\lambda Q|_0^{\infty}}|\Psi\ket
    }\Bigg)^2\quad (X\to\infty).
\end{equation}
We have factorized the expectation value for operators lying far apart from each other; the extensive free-energy part in the large region $[vT,X-vT]$, centered at $X/2$, cancels in the numerator and denominator. This holds assuming appropriate clustering of the state, $|x-y|\langle\Psi|O_1(x)O_2(y)|\Psi\rangle^c\rightarrow0$ as $|x-y|\rightarrow\infty$. We have then used translation invariance and parity symmetry, under which $j$ is odd and $q$ is even, to write
\begin{align}
    &\bra \Psi|e^{\frc\lambda2 Q|_{X/2}^{X}}
    e^{-\lambda J_X|_0^T}
    e^{\frc\lambda2 Q|_{X/2}^{X}}|\Psi\ket\\
    &=\nonumber
    \bra \Psi|e^{\frc\lambda2 Q|_{-X/2}^{0}}
    e^{-\lambda J_0|_0^T}
    e^{\frc\lambda2 Q|_{-X/2}^{0}}|\Psi\ket\\
    &=\nonumber
    \bra \Psi|e^{\frc\lambda2 Q|_{0}^{X/2}}
    e^{\lambda J_0|_0^T}
    e^{\frc\lambda2 Q|_{0}^{X/2}}|\Psi\ket.
\end{align}

The same three steps can be performed in GGEs instead of $|\Psi\ket$. There, one can then show that the result is compatible with the expected non-equilibrium fluctuation relations \cite{Esposito_2009, Bernard_2016, Perfetto_2020} (see App. \ref{FluctuationRelations}), supporting the validity of these steps.

Finally, in a fourth step, if we assume that pairs do not correlate the ray $x=0,t\in[0,T]$ (see Fig.\ref{fig:quench}), one may use the quench steady states locally on each region $x<0$, $x>0$ at $t=0$ (this is a local relaxation assumption, see \cite{LongPaper} for a discussion and a check in a free fermions model). Using \eqref{longtimepsi}, \eqref{longtimepsiQ}, resp., this gives the partitioning protocol (as $T\to\infty$)
\begin{equation}
    \frc{\bra \Psi|e^{\frc\lambda2 Q|_{0}^{\infty}}
    e^{\lambda J_0|_0^T}
    e^{\frc\lambda2 Q|_{0}^{\infty}}|\Psi\ket}{\bra \Psi|e^{\lambda Q|_{0}^{\infty}}|\Psi\ket}
    \asymp
    \frc{\Tr \big[e^{\lambda J_0|_0^T}\, e^{-W_\Psi|_{-\infty}^0
    -W_{\Psi}^\lambda|_0^\infty}\big]}{
    \Tr \big[e^{-W_\Psi|_{-\infty}^0
    -W_{\Psi}^\lambda|_0^\infty}\big]
    }
    \label{mainfactorisation}
\end{equation}
which shows \eqref{fdyn} with \eqref{partitioning}. Formally, this holds assuming strong enough decay of connected correlations for the current operators at different times, $|t-t'|\langle\Psi|e^{\frc\lambda2 Q|_0^\infty}j(0,t)j(0,t')e^{\frc\lambda2 Q|_0^\infty}|\Psi\rangle^c\rightarrow0$ as $|t-t'|\rightarrow\infty$.

\section{Curved trajectories and the violation of temporal clustering}
\label{TClusteringViolationMain}

The fourth step above assumes decay of correlations in time, hence no long-range correlations. There are (at least) two sources of potential long-range correlations: those arising from the inhomogeneous fluid state as shown by the BMFT \cite{Doyon_LongRangeCorrelations}, and those emerging from the quench as discussed above. The former do not affect the current FCS in the partitioning protocol \cite{Doyon_LongRangeCorrelations,doyon2022ballistic}. The latter, according to Fig.~\ref{fig:quench}, also appear not to time-correlate the origin $x=0$.

\begin{figure}
    \centering
    \includegraphics[width=0.8\linewidth]{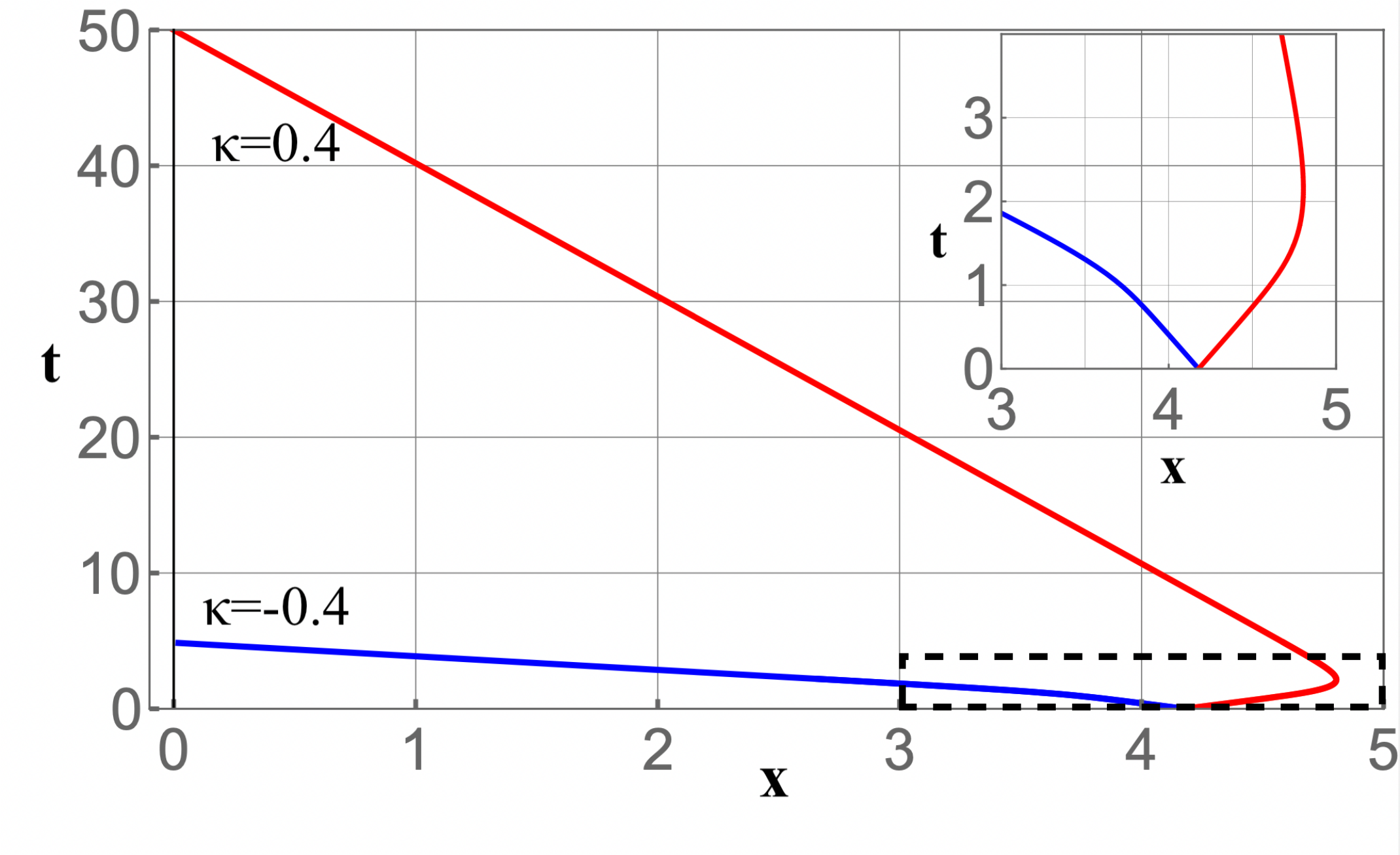}
    \caption{Illustration of curved trajectories of Euler-scale hydrodynamical modes with opposite rapidities after the bi-partitioning protocol with initial density matrix $e^{-W_\Psi|_{-\infty}^0 - (W_\Psi - 2\lambda Q)|_{0}^\infty}$ in the LL model. The inset magnifies and better visualizes the region compassed by black dashed lines. The trajectories were constructed via the method of characteristics using the space-time dependent (effective) velocity field $v^\text{eff}(k,\xi=x/t,\lambda)$. The parameters are $c=1, d=2, \lambda=0.75$ and the rapidities are $\kappa= \pm 0.4$ for the right (red) and left (blue) pairs of trajectories, respectively.}
\label{fig:CurvedTrajectories}
\end{figure}

However, the picture in Fig.~\ref{fig:quench} {\em omits the curvature of hydrodynamic trajectories in inhomogeneous fluid states}, which we call "hydrodynamic waft effect". Emitted at $t=0$ are correlated pairs of fluid modes, which follow fluid characteristics in time. In interacting models, the fluid flow induced by the partitioning protocol gives rise to a hydrodynamic waft which curves characteristics and may make initially right-moving modes into left-moving modes (and vice versa), see Fig.~\ref{fig:CurvedTrajectories}. The impact of such curved trajectories on the evolution of entanglement in certain inhomogeneous cases have been discussed in \cite{CurvedTrajectories2019}. For $\lambda$ small, the bias in the partitioning protocol is small, hence the effect is small. Then, only modes with small hydrodynamic velocities $\propto \lambda$ have the potential to change direction and produce long-range correlations on $x=0$, $t\in[0,T]$. In typical integrable models, there is a continuum of hydrodynamic modes \cite{castroalvaredo2016emergent, bertini2016transport}, parametrized by quasi-momentum $\kappa\in\R$. Modes with small hydrodynamic velocities have small quasi-momentum $|\kappa|< \Lambda\propto \lambda$ (cf.~Fig. \ref{fig:Big}) and produce small currents $\mathsf j \propto \kappa$. Long range correlations will only affect cumulants $c_n^\lambda$, $n\geq 2$ in the current FCS $f_{\rm dyn}=\sum_n c_n^\lambda \lambda^n/n!$, and our power-counting argument gives $c_n^\lambda\propto \mathsf j^n \propto \lambda^n$. As $c_n^\lambda$ occurs at order $\lambda^n$ in the current FCS, the lowest order where corrections may occur in $F(\lambda)$ is $\lambda^{n}\times \lambda^n$ for $n=2$, thus $c_m$ for $m\geq 4$. Further, the density of pairs produced is symmetric under $\kappa\to-\kappa$ as particles are emitted with equal and opposite momenta and by analyticity in $\kappa$, and if quasiparticles have fermionic statistics, Dirac exclusion implies vanishing at $\kappa=0$, hence a density $\sim\kappa^2$. This gives an additional factor $\propto \lambda^2$, thus, in many-body integrable models with fermionic statistics, only cumulants $c_n$ for $n\geq 6$ may be affected. We provide supporting calculations and evidence for this analysis in Sec.\ref{Sec:Curved}.

In integrable models, formula \eqref{main} specializes to a conjecture proposed in \cite{Bertini_2023_tFCS, Bertini_2024_tFCS}, which is therefore here argued not to be generically exact for cumulants beyond order 5. In order to confirm this, using standard integrability techniques, we analyze trajectories in the above partitioning protocol corresponding to the BEC quench of the LL model, as shown in Fig.~\ref{fig:CurvedTrajectories}. 
For finite $\lambda$, curved trajectories exist, producing long-range correlations. See Sec. \ref{Sec:Curved} and App. \ref{App:NumericallyInvestigatedCurvedTraj} for details, and App. \ref{Clustering} for the clustering properties of the BEC state.
Formula \eqref{main} is expected to give {\em all} scaled cumulants $c_n$ in (at least) three situations. First, if the model, integrable or not, {\em admits at most one positive, and one negative hydrodynamic velocity}. Then the quench can only emit correlated pairs of fluid modes traveling in opposite directions, and for every $\lambda$ small enough, trajectories cannot be curved to produce long-range correlations on $t\in[0,T]$ at $x=0$. This includes the rule-54 model, whose FCS was shown to be given by the specialization of \eqref{main} to this model \cite{Bertini_2023_tFCS, Bertini_2024_tFCS}. With two velocities of the same sign, one may deform the contours by slanting the ``vertical branch" (c.f. Fig. \ref{fig:quench}) at a slope lying within the two velocities, again avoiding correlations: thus the formula may be generalized to this case. Second, in {\em free particle models} (or whenever the Euler hydrodynamic equation is linear): then trajectories do not curve. Finally, if the initial state is an eigenstate of $Q$: then $e^{\frc \lambda 2 Q}|\Psi\ket \propto |\Psi\ket$ and there is no partitioning, so the fluid state is uniform hence trajectories are straight.

\section{The hydrodynamic flow equations for $f_\text{dyn}(\lambda)$ in integrable models}
\label{Sec:TBA}

In this section we show how the main formula \eqref{fdyn} for $f_\text{dyn}(\lambda)$ can be evaluated in integrable quenches. The evaluation requires the solution of the BFT flow equations, which we explicitly write down. For the sake of brevity and transparency, we shall specify the flow equations for the Lieb-Liniger model, whose thermodynamic Bethe Ansatz treatment is reasonably simple, but the formula is easy to generalise to other integrable models. With the Lieb-Liniger model at hand, we shall focus on the specific example of an integrable quench where the initial state is a Bose--Einstein condensate (BEC).

\subsection{Describing macro-states in integrable models}
Integrable models admit an extensive number of (quasi-) local commuting conserved quantities charges  $Q_{l},~l=1,2,\dots$ including the Hamiltonian. Consequently such models have a stable set of quasi-particle excitations which are indexed by a discrete species index $n=1,\dots,N_s$ and parameterized by a continuous rapidity $k^{(n)}_j$, $j=1,\dots,M_{n}$, but simplicity, we restrict our discussion to models with only one species. An eigenstate of the model is specified by the rapidities of quasi-particles, and is denoted by $|\bm{k}\rangle= |k_1,\ldots,k_{M}\rangle$.
These states are simultaneous eigenstates of all the conserved charges, namely $Q_l |\bm{k}\rangle = \sum_{j=1}^{M} q_l(k_j)|\bm{k}\rangle$.
Conserved charges can be categorized as even or odd ones wrt. spatial parity such as the particle number $N=Q_0$ or the Hamiltonian $H=Q_2$; and  the total momentum $P=Q_1$, respectively. For simplicity we denote the charge, momentum and energy of a quasi-particle of rapidity $k$ by $q (k)=q_0(k),p(k)=q_1(k)$ and $E(k)=q_2(k)$.   
In the thermodynamic limit and at finite (particle, energy, etc) density the model can be treated using the methods of the thermodynamic Bethe Ansatz (TBA)~\cite{takahashi1999thermodynamics}. Such states are described through distributions related to their quasi-particle content, in particular, $\rho(k),\rho^h(k),\rho^t(k)$ and $\theta(k)$ which are respectively the distribution of occupied, the distribution unoccupied quasiparticles,  the total density of states $\rho^t(k)=\rho(k)+\rho^h(k)$ and the occupation function $\theta(k)=\rho(k)/\rho^t(k)$.  These functions are related to each other by the Bethe-Takahashi equations 
\begin{eqnarray}
\rho^t(k)=\frac{p'(k)}{2\pi}+\varphi\star\rho(k)
\end{eqnarray}
where $(\cdot)'$ denotes differentiation wrt. $k$ and $\star$ is the convolution $f\star g(x)=\int \frac{{\rm d}y}{2\pi} f(x-y)g(y)$. $\varphi(k-k')$ is the scattering kernel which characterizes the scattering between quasi-particles with rapidities $k$ and $k'$.  

When macro-states are considered containing  many excitations the bare quasi-particle properties become dressed due to the interactions. Dressed quantities denoted by $(\cdot)^{\rm dr}$ satisfy the  integral equation
\begin{equation}
f^{\text{dr}}(k)=f(k)+\left(\varphi\star \theta\,f^{\text{dr}}\right)(k)\,.\label{eq:Dressing}
\end{equation}
and
\begin{equation}
v^{\text{eff}}(k)=\frac{\left(E'(k)\right)^{\text{dr}}}{\left(p'(k)\right)^{\text{dr}}}\,,
\label{vEff}
\end{equation}
where $E'$ and $p'$ prime are the derivatives if the 1-particle energy and momentum. 
To better expose the flow equations for $f_\text{dyn}(\lambda_i)$ we specified an important quantum model, the Lieb--Liniger model defined  by
\begin{equation}\label{eq:Hamiltonian}
H_\text{LL}\!=\!\frac{1}{2}\int \!\!{\rm d}x\, b^\dag(x)\!\!\left[-\frac{\partial_x^2}{m}\right]\!b(x)+2c\, b^\dag(x)b(x)b^\dag(x)b(x).
\end{equation}
where we will set $m=1$ and $b^\dag(x),b(x)$ are canonical spinless  bosonic operators satisfying $[b(x),b^\dag(y)]=\delta(x-y)$. The Hamiltonian \eqref{eq:Hamiltonian} describes bosons of mass $m$ interacting via a density-density interaction of strength $c>0$ corresponding to repulsion between the bosons. For the out-of-equilibrium dynamics certain integrable quenches are considered \cite{piroli2017what}. Via the Quench Action method \cite{caux2013time, caux2016quench}, it is possible to characterize the steady-state GGE for such quenches. These quenches also have the distinctive property that the initial states have non-vanishing overlaps with the eigenstates of the post-quench Hamiltonian only if these states consist of pairs of particles with opposite momentum/rapidity.
For the LL model we consider the Bose-Einstein condensate states as such integrable initial state (BEC) defined as
\beq
|\Psi\rangle_\text{BEC}=\lim_{N,L\rightarrow \infty}\frac{(b^\dag_{0,L})^N}{\sqrt{N!}}|0\rangle\,,
\label{eq:Psi0LL}
\eeq
respectively,
where $b^\dag_{0,L}$ creates a zero momentum boson in finite volume $L$ and when taking the thermodynamic limit the density $N/L=d$ is kept fixed and the state is also the ground state of the LL model at $c=0$. To see the clustering property of the state, consult App. \ref{Clustering}.

For these initial states it possible to compute the the steady-state spectral densities corresponding to the GGE via solving
\beq
\ln \eta(k)=g(k) -\left(\varphi
\star\ln[1+\eta^{-1}]\right)(k)
\label{QA-GGE}
\eeq
where $\varphi(k)= \frac{2|c|}{c^2+k^2}$ with $g=\ln[k^2(k^2+(c/2)^2)]-2\ln[d\, c]$ for the BEC quench \cite{denardis2014solution}, and where $\theta(k)=1/(1+\eta(k))$.

\subsection{Hydrodynamic predictions for FCS and SCGF}
In the following we specify the results for the FCS after quenches which have been first obtained in \cite{Bertini_2023_tFCS,Bertini_2024_tFCS} via space-time swap, and in this letter using hydrodynamic principles. The validity of these principles in the particularly important case of free models will be presented elsewhere \cite{LongPaper}.  
In particular, applying the description of integrable systems, the predictions of \eqref{main} for the function $f_\text{dyn}(\lambda)$ can be further specified as follows. We write $2f_\text{dyn}(\lambda)=f_{\text{j}_\text{L}}^{\rho_\text{L}}(\lambda)+f_{\text{j}_\text{R}}^{\rho_\text{R}}(\lambda)$, which are associated with the fluctuations of the corresponding current along the left (L) and right (R) regions, c.f. Fig. \ref{fig:quench}. Using the shorthand y=L,R  \eqref{main} dictates that $f_{\text{j}_{y}}^{\rho_\text{y}}(\lambda)$ are defined through the flow equations
\beq
\label{Flow1}
f_{\text{j}_{y}}^{\rho_\text{y}}(\lambda)=\int_0^{\pm\lambda} \!\!\!\!\!\dd \beta\!\!\int \frac{\dd k}{2\pi}E'(k)\tilde{\theta}_\text{y}^{\beta}(k)q^\text{dr}[\tilde{\theta}_\text{y}^{\beta}](k)\,,
\eeq
with the quantities in the integrand satisfying
\beq
\label{Flow2}
\begin{split}
&\partial_{\beta}\ep^{\beta}_\text{y}(k)=-\text{sgn}(v^\text{eff}[\tilde{\theta}_\text{y}^{\beta}](k))q^\text{dr}[\tilde{\theta}_\text{y}^{\beta}](k)\\
&\tilde{\theta}_\text{y}^{\beta}(k)=\frac{1}{\exp [\ep^{\beta}_\text{y}(k)]+1},\quad \text{with} \quad
\tilde{\theta}^{0}_\text{y}(k)=\theta_\text{y}(k)\,.
\end{split}
\eeq
In the equations above we used the dressing operation \eqref{eq:Dressing} and the expression for the effective velocity \eqref{vEff}
in which the rapidity derivatives $E'$ and $p'$ prime are the derivatives of the 1-particle energy and momentum,  are just $2 k$ and $1$ for the LL model, respectively. 

The initial condition for \eqref{Flow2} is specified by the NESS ($x\ll t$) of a bi-partitioning protocol as
\beq
\begin{split}
\label{IC}
\theta_\text{L}(k)&=\Theta_\text{H}(v^\text{eff}(k))\theta(k)+\Theta_\text{H}(-v^\text{eff}(k))\theta^{2\lambda}(k)\\
\theta_\text{R}(k)&=\Theta_\text{H}(v^\text{eff}(k))\theta^{2\lambda}(k)+\Theta_\text{H}(-v^\text{eff}(k))\theta(k)\,,
\end{split}
\eeq
where $\Theta_\text{H}$ is the Heaviside-theta function and the filling function $\theta(k)$ correspond to the GGE $\rho_\text{GGE}$ and $\theta^{2\lambda}(k)$ to a modified GGE defined by the steady state for $\langle \Psi |e^{\lambda/2\, Q}O(t,x)e^{\lambda/2\, Q} |\Psi\rangle$ at late times. For integrable initial states, $\theta^{2\lambda}(k)$ is obtained from Eq. \eqref{QA-GGE} upon replacing $g(k)$ by $g(k)-2\lambda q(k)$ corresponding to $\rho_\text{GGE}e^{2\lambda Q}$.

\section{Curved trajectories and long-range correlations for the BEC quench in the LL model} 
\label{Sec:Curved}

This section presents a detailed analysis of the emergence of curved hydrodynamic trajectories in the Euler-scaling regime of the LL model after the biased BEC quench, and of the enhanced current-current autocorrelations generated by such trajectories. Although our discussion focuses on the LL-BEC setup, the underlying mechanism is expected to hold in a broad class of interacting integrable models.

In Subsec. \ref{Sec_Curved_A}, we study the Euler characteristics associated with the modified bipartite quench problem in interacting integrable models. We show that characteristics corresponding to opposite small quasi-particle momenta become curved, as illustrated in Fig. \ref{fig:CurvedTrajectories}. Moreover, the spectral region supporting such curved trajectories shrinks smoothly as the counting field $\lambda$, controlling the imbalance in the bipartite quench, approaches zero.

In Subsec. \ref{Sec:BMFT}, we then show that these curved trajectories generate enhanced dynamical current correlations after the biased BEC quench. Our analysis is based on a BMFT-inspired picture \cite{doyon2022ballistic}, where initial-state correlations propagate ballistically and noiselessly on top of a locally equilibrated GHD background.
To implement this program, we first need the appropriate initial correlations. 
While these could in principle be studied directly in the LL-BEC setup, we instead compute the analogous quantities in free-fermion integrable initial states, where exact calculations are possible. The justification is that integrable initial states generically exhibit similar large-scale correlation structures across integrable models, due to the pair structure of quasi-particle excitations. As a result, the free-fermion correlations already capture the essential features needed to estimate the emergent dynamical correlations in the LL model. 
Finally, using the Euler characteristics of the LL-BEC problem, we show that for small counting field $\lambda$, the propagation of correlations along the curved trajectories can be tracked explicitly, allowing the resulting dynamical current correlations to be evaluated.

\subsection{Existence of curved hydrodynamic trajectories from GHD}
\label{Sec_Curved_A}
Below we report a detailed numerical analysis to demonstrate the existence of curved hydrodynamic trajectories in the bi-partitioning protocol, that can give rise to long-range correlation for the current-current 2pt function. The analysis is similar to that of \cite{CurvedTrajectories2019} and for simplicity, we shall consider the case of the initial density matrix $\rho_\text{GGE}^{(l)}\otimes \rho_\text{GGE}^{(r)}e^{2\lambda Q}$. Let us now formulate precisely what we want to demonstrate:
$\forall\, c>0,d>0,\lambda>0\,, \exists\,\Lambda>0 $ such that $\forall \kappa \in [-\Lambda, \Lambda]$ and $\forall T>0 \quad \exists \,T'>0$ such that
\beq
\begin{split}
&X(T,\kappa)=0 \quad\text{and}\quad X(0,\kappa)=x_0\,,\quad \text{and}\\
&X(0,-\kappa)=0\quad \text{and} \quad X(0,-\kappa)=x_0\,,
\end{split}
\label{CritForWafting}
\eeq
for some real $x_0$, where $X(t,\pm\kappa)$ are  solutions of the Euler characteristic equation, i.e.,
\beq
\dot{X}(t,\kappa,\lambda)=v^\text{eff}(\kappa,X/t,\lambda) \quad \text{and}\quad X(0,\kappa,\lambda)=x_0
\label{DiffEq}
\eeq
where we remind ourselves that $\xi=x/t$ and $v^\text{eff}(\kappa, \xi, \lambda)$ is the effective in the $\xi$-dependent GGE emerging after the bi-partite quench with the aforementioned initial density matrix.

We note that the computation of the ray-dependent effective velocity following a bi-partitioning protocol is a standard problem in GHD. For its identification we first recall that the effective velocity in GGEs can be obtained by the dressing equations 

The quantity $\theta$ is the filling function of the GGE, which has to be determined a self-consistent way via
\begin{equation}
\begin{split}
&\theta(\kappa,\xi,\lambda)=\\
&\Theta_{H}(v^{\text{eff}}(\kappa,\xi,\lambda)-\xi)\theta_\text{L}(\kappa)+\Theta_{H}(\xi-v^{\text{eff}}(\kappa, \xi,\lambda))\theta_\text{R}(\kappa)\,,
\end{split}
\label{GHDBasicEq1XiSolution}
\end{equation}
where $\theta(\kappa)_\text{L/R}$ are the filling functions of the GGEs characterizing the initial system on its left/right parts, i.e., corresponding to $\rho_\text{GGE}$ of the steady-state or $\rho_\text{GGE}e^{2\lambda Q}$.
The above equation can be rewritten also as
\begin{equation}
\begin{split}
&\theta(\kappa,\tau,\lambda)=\Theta_{H}(\kappa-\tau)\theta_\text{L}(\kappa)+\Theta_{H}(\tau-\kappa)(\kappa))\theta_\text{R}(\kappa)\,,\, \text{with}\\
&\xi=v^\text{eff}[\theta(\kappa,\tau,\lambda)](\kappa=\tau)\,,
\end{split}
\label{GHDBasicEq1Theta0iSolution}
\end{equation}
which is an implicit equation for $\tau$ if the the value of the ray $\xi$ is prescribed.
Once the ray-dependent effective velocity is numerically computed, it is immediate the reconstruct the $x,t$-dependent effective velocity field $v^\text{eff}(\kappa, x,t,\lambda)=v^\text{eff}(\kappa, x/t,\lambda)$ via which the equation \eqref{DiffEq} can be solved numerically.
Although the numerical solution of these equations may not justify the criterion \eqref{CritForWafting}, it is instructive to study their fulfillment, which is indeed the case as demonstrated by Fig. \ref{fig:CurvedTrajectories}. Showing that \eqref{CritForWafting} holds can be demonstrated without eventually solving the differential equation \eqref{DiffEq} in the following way, where we shall make use of numerical observations while carefully exploring the parameter space. 

\begin{figure}
    \centering
    \begin{subfigure}
    \centering
     \includegraphics[width=0.48\linewidth]{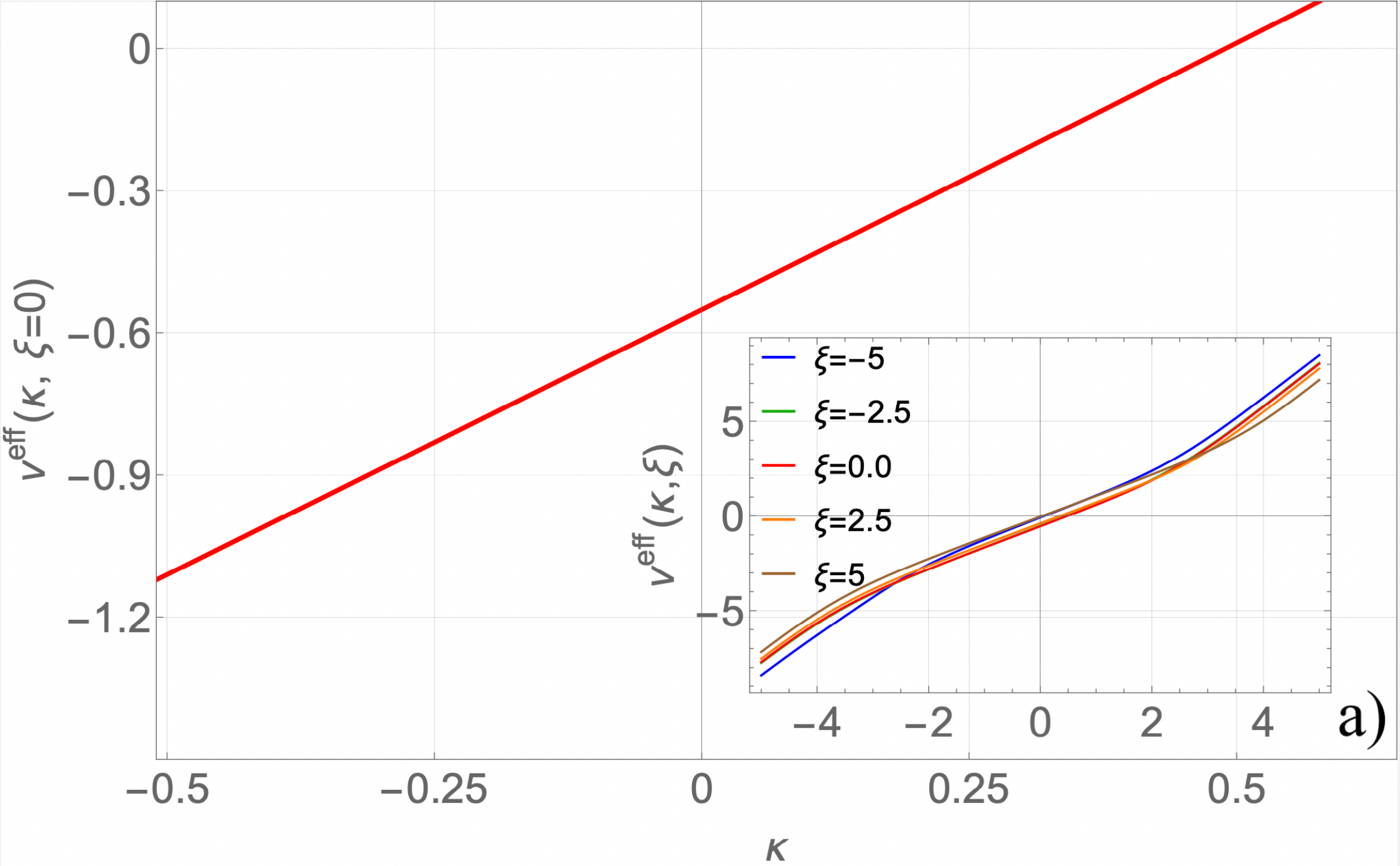}
    \end{subfigure}
    ~
    \begin{subfigure}
    \centering
     \includegraphics[width=0.48\linewidth]{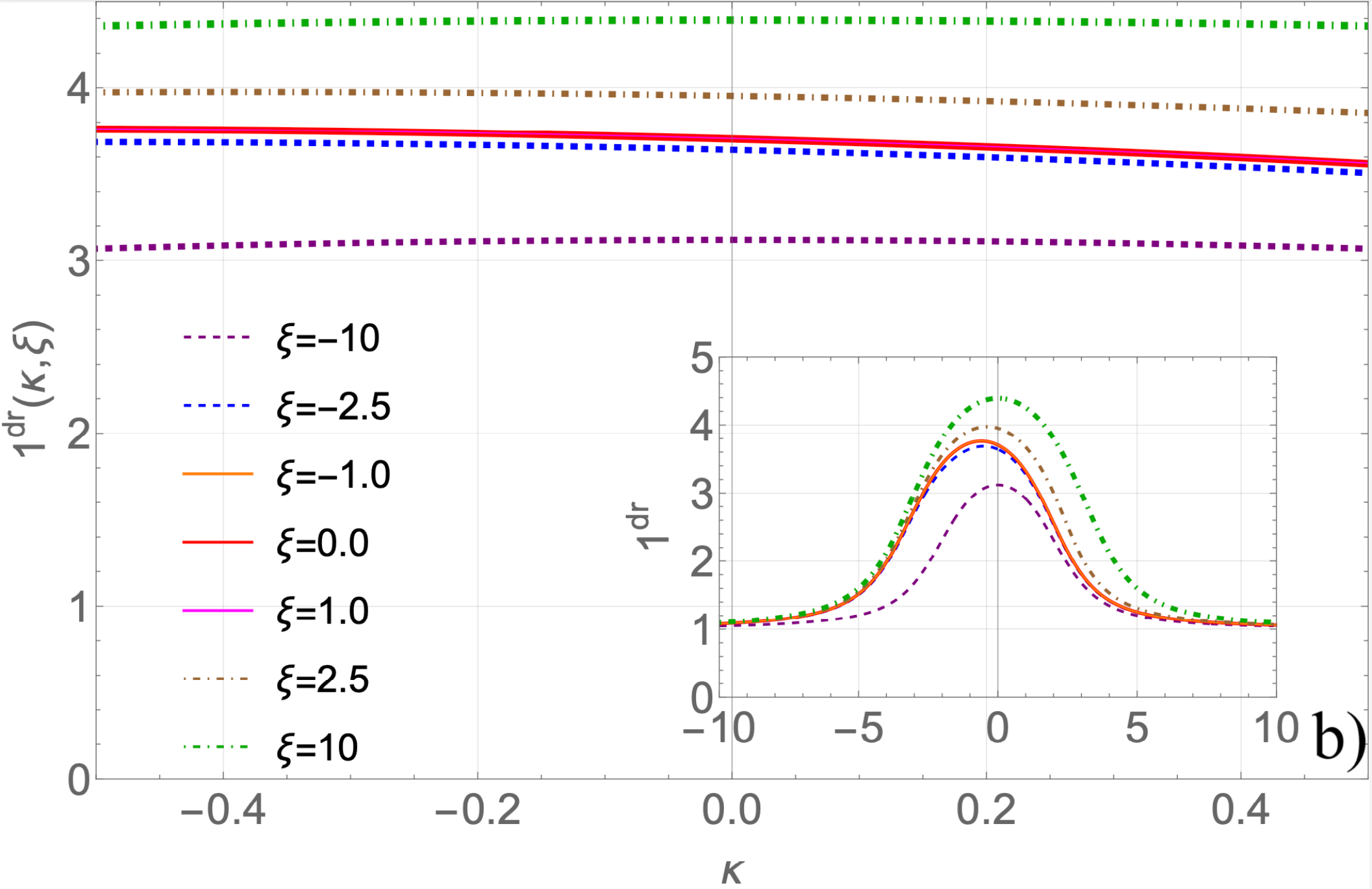}
    \end{subfigure}\\
     \begin{subfigure}
    \centering
     \includegraphics[width=0.48\linewidth]{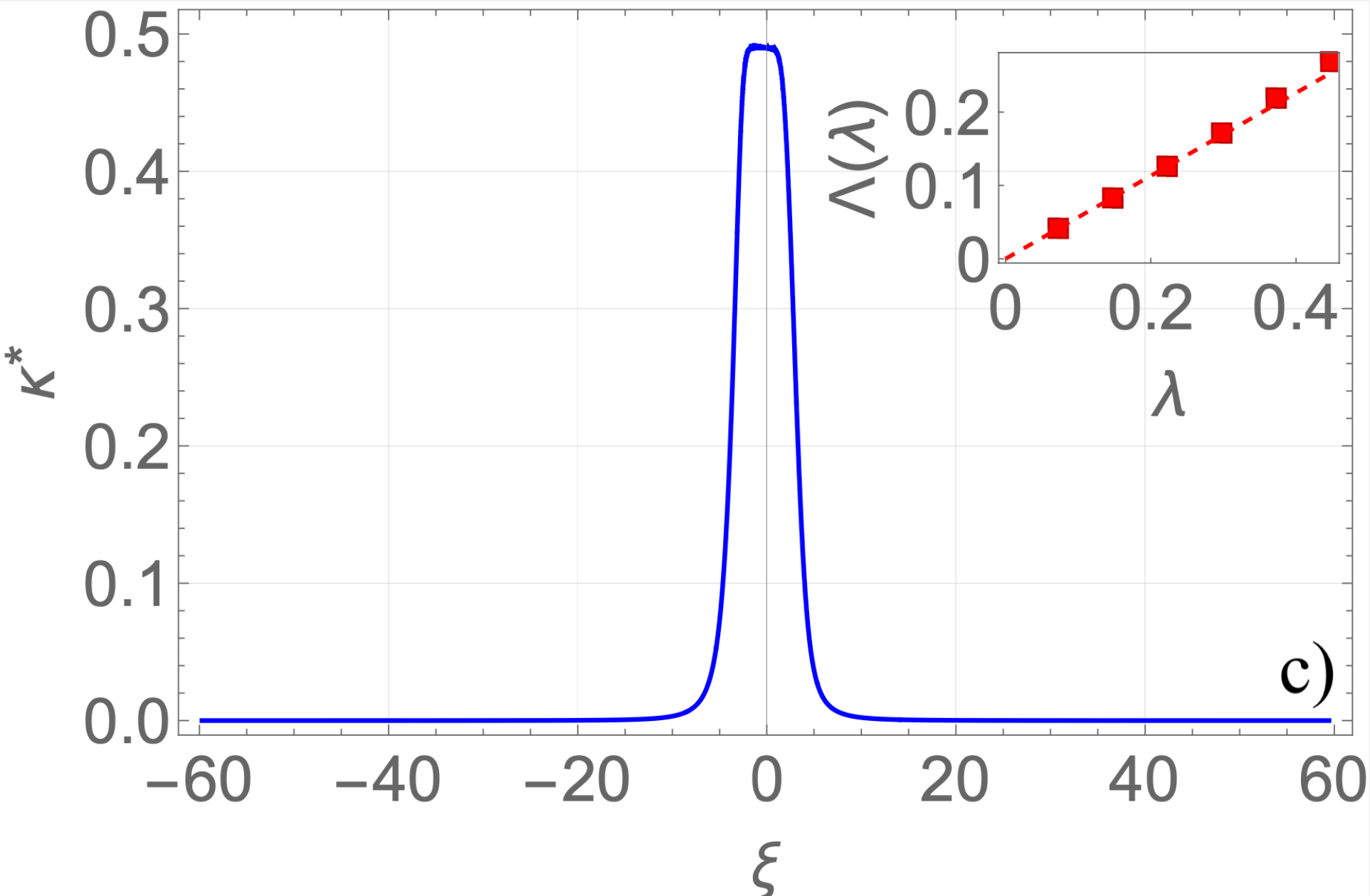}
    \end{subfigure}
    ~
    \begin{subfigure}
    \centering
     \includegraphics[width=0.48\linewidth]{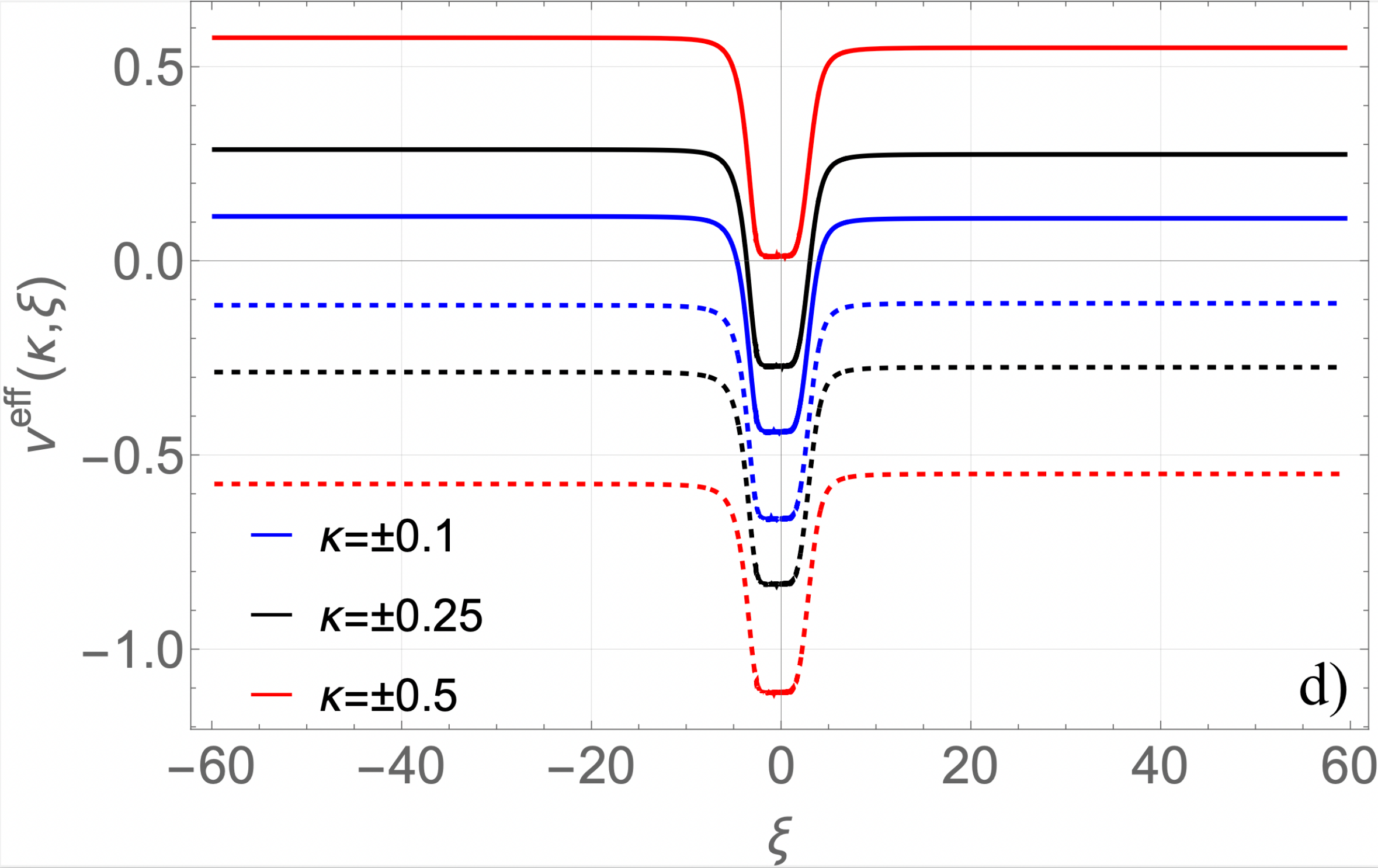}
    \end{subfigure}\\
    
    \caption{a) The effective velocity in the NESS ($\xi=0$) for $\kappa \in [-\Lambda,\Lambda]$ with $\Lambda\approx0.5$, and the effective velocities for other ray-values (inset), $\xi=-5,-2.5,0,2.5,5$ corresponding to colors blue, red, black, orange and brown. b) $1^\text{dr}(\kappa,\xi)$ for $\kappa \in [-\Lambda,\Lambda]$ in the NESS and for further $\xi$-values, as well as at larger $\kappa$ (inset), $\xi=-10,-2.5,-1,0,1,2.5,10$ corresponding to purple (dashed), blue (dashed), orange, red, magenta, brown (dotted-dashed) and green (dotted-dashed) lines, respectively. c) The zero of $v^\text{eff}(\kappa^*,\xi)$ for fixed $\xi$ as function of $\xi$ and the linear dependence of $\Lambda=\kappa^*(\xi=0)$ on $\lambda$ (inset). d) The effective velocity $v^\text{eff}(\kappa,\xi)$ as a function of $\xi$ for $\pm\kappa$ values from $[-\Lambda,\Lambda]$, where the continuous and dashed lines correspond to positive and negative rapidities, respectively, blue $\xi=0.1$, black $\xi=0.25$ and red $\xi=0.5$. The parameters are $c=1$, $d=2$ and $\lambda=0.75$.}
    \label{fig:Big}
\end{figure}

The first observation is that for $\forall\, c>0,d>0,\lambda>0$, it is true that for any fixed ray $\xi$, $v^\text{eff}(\kappa,\xi,\lambda)=0$ has only one unique solution $\kappa^*(\xi)$. In the particular geometry ($\rho_\text{GGE}$ on the left, and $\rho_\text{GGE}e^{2\lambda Q}$ on the right side of the system) this $\kappa^*$ value is non-negative and lies in the interval $[0,\tilde{\Lambda}]$ and $\kappa^*(\xi)$ reaches its maximum value around $\xi\approx 0$ and $\kappa^*\approx 0$ for large enough $\xi$-s which, for smaller interaction strength and densities are close to zero, c.f. Fig \ref{fig:Big}
c). These facts imply that first of all, far away from the NESS,
\beq
\sgn(v^\text{eff}(\kappa, \xi,\lambda))=\sgn(\kappa)\quad \text{when}\quad x\gg t
\label{VeffXi1}
\eeq
that is, the sign of the effective velocity of hydrodynamics modes with opposite rapidities is different.
The other implication is that close to the  NESS, i.e., $\xi\approx 0$ or $t \gg x$ this is not the case; instead for 
\beq
\forall \kappa \in [-\Lambda, \Lambda],\quad \sgn(v^\text{eff}(\kappa, \xi,\lambda))=-1\quad\text{when}\quad t\gg x
\label{VeffXi2}
\eeq
(c.f.  Fig. \ref{fig:Big} a)) where $\Lambda\approx \tilde{\Lambda}$, $\Lambda\propto \lambda$ as reported by Fig.  \ref{fig:Big} c) and the sign of the velocity is $-1$ in the particular quench geometry.
The final, crucial observation is that whereas  $\forall \kappa \in[-\Lambda,0]$, $v^\text{eff}(\kappa, \xi,\lambda)$ is negative for any $\xi$, for $\forall \kappa \in[0,\Lambda]$, $v^\text{eff}(\kappa, \xi,\lambda)$ can be both positive ($|\xi|\gg 1$, \eqref{VeffXi1}) and negative ($|\xi|\ll 1$, \eqref{VeffXi2}) and in particular it has two zeros
$\xi^*_1<0\,\text{ and }\,\xi^*_2>0$ (satisfying
$v^\text{eff}(\kappa, \xi^*_{1,2})=0$), c.f.  Fig. \ref{fig:Big} d). Furthermore, it is true that 
\beq
\begin{split}
&\forall \kappa \in[0,\Lambda],\quad |\max_\xi v^\text{eff}(\kappa, \xi,\lambda)|,  |\min_\xi v^\text{eff}(\kappa, \xi,\lambda)|<|\xi^*_1|\\
&\qquad\quad \text{and}\quad\,\,\, |\max_\xi v^\text{eff}(\kappa, \xi,\lambda)|,  |\min_\xi v^\text{eff}(\kappa, \xi,\lambda)|<\xi^*_2\,,
\label{VeffXi3}
\end{split}
\eeq
as visualized by Fig. \ref{fig:Big} d).
From the above, it follows that any hydrodynamic trajectory $X(t,\kappa,\lambda)$ with $\kappa \in[0,\Lambda]$ and with initial condition $X(0,\kappa,\lambda)=x_0>0$ over the course of its time evolution has initially positive velocity, however, at some finite time the velocity changes sign to negative. After this instant the velocity cannot change sign anymore (as $|\min_\xi v^\text{eff}(\kappa, \xi,\lambda)|<|\xi^*_1|, \xi^*_2$), therefore the trajectory with initial position $x_0>0$ will necessarily cross the $x=0$ point. Additionally, another trajectory with $X(0,\kappa,\lambda)=x_0, \forall \kappa \in [-\Lambda,0]$ has only negative velocities and hence also necessarily reach the $x=0$ value. That is, taking two trajectories with opposite rapidity $\kappa,-\kappa \in [-\Lambda,\Lambda]$ with the same initial position $x_0>0$ both trajectories will necessarily cross the $x=0$ point. Finally, because the trajectories $X(t,\kappa>0,\lambda)$ will initially have positive then negative velocities irrespective of the initial position $x_0>0$, and this velocity is bounded (c.f. Fig. \ref{fig:Big} d)), it also follows that upon $x_0\rightarrow \infty$, $T\rightarrow \infty$ as well ( where $X(T,\kappa,\lambda)=0$). In other words, the curved trajectories are not only present in a finite space-time region, but $T$ can be arbitrarily large, which is an important requirement for the onset of the dynamical long-range correlations. For further numerical analysis, and for the precise range of the investigated parameter space consult App. \ref{App:NumericallyInvestigatedCurvedTraj}.
Additionally,  in Appendix \ref{FirstOrderGHD} by expanding the GHD equations with respect to the counting field t first order, we demonstrate that the onset of curved trajectories occurs continuously and does not appear only above a critical nonzero value of $\lambda$. 

\subsection{Dynamical long-range correlations}
\label{Sec:BMFT}

One main finding of this work is that curved hydrodynamic trajectories together with spatial short-range initial correlation and certain initial correlations in momentum or rapidity space give rise to long-range correlations for the time integrated current, hence the predictions of BFT for the initial time evolution of the scaled cumulants of conserved charges are not correct. More precisely, the connected correlation function 
\beq
\frac{\langle\Psi|e^{\lambda/2 \,Q|_0^\infty(0)}j(0,t)j(0,t')e^{\lambda/2 \,Q|_0^\infty(0)}|\Psi\rangle^c}{\langle\Psi|e^{\lambda Q|_0^\infty(0)}|\Psi\rangle}
\eeq
at finite counting field, as well as higher point connected correlation functions exhibit slow decay upon separating the time arguments of the currents, which prevents the applicability of BFT to compute the FCS of the current after the bipartite state from $e^{\lambda/2\, Q|_0^\infty(0)}|\Psi\rangle$,  which translates into the time evolution of the FCS of the corresponding integrated conserved charge. 

In the following we demonstrate the validity of the above claim via the example of the LL model and the BEC quench using a series of non-trivial arguments.
Therein, a particularly important role is played by the BMFT whose main assumption is that spatio-temporal fluctuations, and hence also correlations  can be described by considering the fluctuations in the initial state and appropriately transporting them via the Euler hydrodynamic equations of the conserved quantities in integrable systems admitting ballistic transport. 

The other cornerstone regards the correlations in the initial state. In particular, since in the post-quench expansion of integrable initial state excitations with opposite momentum or rapidity are correlated we expect the core correlations
to be those among rapidity-resolved charges or hydrodynamical normal mode.  While the intuition for this claim is natural, the characterization of such correlations via rapidity-resolved conserved quantities in genuinely interacting integrable models is non-trivial. 

Regarding the construction of rapidity-resolved conserved quantities in interacting integrable models, this task has been achieved in \cite{Ilievski_2016,Ilievski_2017,doyon2023abinitioderivationgeneralised}. However, for our purposes and to apply the BMFT equations, the (semi-)locality properties of such charges are an important ingredient, which, however, have only partially been studied and verified in \cite{Ilievski_2017,doyon2023abinitioderivationgeneralised}. On the contrary, the construction of rapidity-resolved conserved quantities is particularly simple in free fermion models and their semi-locality properties are also well-established \cite{del_Vecchio_del_Vecchio_2024}. 

In the following, therefore, we shall refer to free fermion quenches which is also motivated by one more technical reason. Whereas computing the rapidity-resolved correlations of conserved charges after the BEC quench is possible, at least in principle, the computation would require a linked-cluster expansion with a three-fold summation over LL eigenstates using exact overlap formulas and the form factors of the the  model, and hence might not be practically feasible.
To avoid such technicalities we rather consider the rapidity-resolved correlation functions in the free fermion model after free fermion integrable quenches, exploiting the key structural similarity of integrable quenches, i.e., consisting of pairs of excitations with opposite momentum. In other words, we shall infer the input for the BMFT equations for the LL model by first studying the correlation functions of rapidity resolved conserved charges in the free fermion model.

\subsubsection{Rapidity-resolved conserved quantities}

As we have mentioned there are known constructions for rapidity-resolved conserved charges in interacting integrable models, however, for various reasons we shall consider the corresponding charges in the free-fermion model, where the construction is particularly straightforward \cite{del_Vecchio_del_Vecchio_2024}. 
Using the standard Fourier decomposition of the fermion field operators, 
\beq
\begin{split}
\psi(x,t) &= \frc1{\sqrt{2\pi}}\int \dd k\,e^{\ri k x - \ri E_k t} a_k\\
\psi^\dagger(x,t)& = \frc1{\sqrt{2\pi}}\int \dd k\,e^{-\ri k x +\ri E_k t} a^\dagger_k\,,
\end{split}
\eeq
and
\begin{equation}
a_k=\frac{1}{\sqrt{2\pi}}\int \text{d}k e^{-\ri kx}\psi(x)\,,\quad
a^{\dagger}_k=\frac{1}{\sqrt{2\pi}}\int \text{d}k e^{\ri kx}\psi^{\dagger}(x)
\end{equation}
(where we omitted the trivial time dependence)  with basic anti-commutators
\beq
	\{\psi^\dag(x),\psi(y)\} = \delta(x-y),\quad
	\{a_k^\dag,a_p\} = \delta(k-p).
\eeq
where we can write a conserved charge of the particle number as \cite{del_Vecchio_del_Vecchio_2024}
\beq
Q=\sum_{\kappa_n}Q_{\kappa_n}\,, \text{ with }\quad Q_{\kappa_n}=\int_{\kappa_n-\epsilon/2}^{\kappa_n+\epsilon/2} \!\!\!\!\!\dd k\, a^\dagger_k a_k
\eeq
where $\kappa_n=n \epsilon$. The rapidity-resolved particle density can be written as well \cite{del_Vecchio_del_Vecchio_2024} as
\beq
q(x,t)=\sum_{\kappa_n}q_{\kappa_n}(x,t)\,,
\eeq
with
\beq
\begin{split}
q_{\kappa_n}(x,t)=&\frac{1}{2\pi}\int_{-\infty}^\infty \!\!\!\dd k\, \dd k'\, e^{-\ri (k-k')x}e^{\ri (E_k-E_{k'})t}\times\\
&\quad\times\Theta_\text{H}\left(\frac{\epsilon-|k+k'-2\kappa_n|}{2}\right)a^\dagger_k a_{k'}\,,
\end{split}
\eeq
where $\Theta_\text{H}$ is the Heaviside Theta function and more explicitly
\beq
\begin{split}
&\Theta_\text{H}\left(\frac{\epsilon-|k+k'-2\kappa_n|}{2}\right)=\\
&\Theta_\text{H}\left(\frac{\epsilon-(k+k')+2\kappa_n}{2}\right)-\Theta_\text{H}\left(\frac{-\epsilon-(k+k')+2\kappa_n}{2}\right)\,.
\end{split}
\label{HeavisideThetas}
\eeq

\subsubsection{Rapidity-resolved spatial correlations in the fermionic initial states}

As we have outlined before, a crucial ingredient to evaluate dynamical long-range correlations in the LL model, is the knowledge of the rapidity-resolved spatial correlation of the particle number operator at the initial state. More precisely, however, these initial correlations must be regarded already at the Euler scale, due the requirements of BMFT, that is, we are interested in the quantity
\beq
\begin{split}
&\lim_{t\rightarrow 0}\lim_{\ell\rightarrow\infty}\ell\times_\text{BEC}\!\langle\Psi|q_{\kappa_n}(\ell x,\ell t)q_{\kappa_{n'}}(\ell x',\ell t)|\Psi\rangle^c_\text{BEC}\\
&\quad\neq\lim_{\ell\rightarrow\infty}\ell\times_\text{BEC}\!\langle\Psi|q_{\kappa_n}(\ell x,0)q_{\kappa_{n'}}(\ell x',0)|\Psi\rangle^c_\text{BEC}\,.
\end{split}
\eeq
The computation of this object for the LL model is notoriously complicated and, in principle, might be carried out using a non-trivial form factor expansion and resummation. For this reason, as already explained, we rather compute the corresponding 2pt function in fermionic initial states and will use the result as initial input for the BMFT equations for the interacting case. This approximation is justified by the universal structure of integrable initial states, which are made up solely by correlated pairs of excitations with opposite momentum.
Therefore, we first consider
\beq
\lim_{t\rightarrow 0}\lim_{\ell\rightarrow\infty}\ell\times_\text{FF}\!\langle\Psi|q_{\kappa_n}(\ell x,\ell t)q_{\kappa_{n'}}(\ell x',\ell t)|\Psi\rangle^c_\text{FF}\,,
\eeq
where $|\Psi\rangle_\text{FF}$ is characterized by a  K-function $K(k)$ or by $K(k)e^{2\lambda}$ as
\beq
|\Psi\rangle_\text{FF}=\mathcal{N}_\text{FF}^{-1/2}\left(\exp{\int_0^\infty\frac{\dd k}{2\pi}K(k)a^\dagger_k a^\dagger_{-k}}\right)|0\rangle\,.
\label{IntISFF}
\eeq
Importantly, the free fermion quench problem can be analyzed exploiting the Bogolyubov transformation
\beq
a_p=u_p\tilde{a}_p+v_p\tilde{a}^\dagger_{-p}\,,\qquad
a^\dagger_p=v^*_{p}\tilde{a}_{-p}+u_p\tilde{a}^\dagger_{p}\,,
\eeq
where
\beq
u_p=\frac{1}{\sqrt{1+|K(p)|^2}}\,,\, v_p=\frac{K(p)}{\sqrt{1+|K(p)|^2}}\,
\eeq
with the new fermion operators $\tilde{a}_p$ annihilating $|\Psi\rangle_\text{FF}$.
\pagebreak
\begin{widetext}
Using this transformation, the 2pt density correlation function is easy to express \cite{LongPaper} from which, using \cite{del_Vecchio_del_Vecchio_2024}, we obtain the momentum-resolved correlation function as
\beq
\begin{split}
&\phantom{}_\text{FF}\langle\Psi|q_{\kappa_n}(\ell x,\ell t)q_{\kappa_{n'}}(\ell x',\ell t)|\Psi\rangle^c_\text{FF}=\int \frac{\dd k}{2\pi}\frac{\dd p}{2\pi}e^{-\ri(k-p)(x-x')}e^{2\ri  t(E_k-E_p)}\Theta_\text{H}\left(\frac{\epsilon-|k+p-2\kappa_n|}{2}\right)\times\\
&\qquad\qquad\qquad\qquad\qquad\qquad\qquad\quad\times \left\{ \mathcal{K}_1(k,p)\Theta_\text{H}\left(\frac{\epsilon-|k+p-2\kappa'_n|}{2}\right)+\mathcal{K}_2(k,p)\Theta_\text{H}\left(\frac{\epsilon-|k+p+2\kappa'_n|}{2}\right) \right\}\,,
\end{split}
\label{ThetaCorrel1}
\eeq
(for brevity we did not write down the same formula with $K(k)e^{2\lambda}$), and we introduced the shorthand notations
\beq
\mathcal{K}_1(k,p)=\frac{|K(k)|^2}{1+|K(k)|^2}\frac{1}{1+|K(p)|^2}\,,\quad
\mathcal{K}_2(k,p)=\frac{K^*(k)}{1+|K(k)|^2}\frac{K(p)}{1+|K(p)|^2}\,.
\label{calK}
\eeq
In the following we show that the last line gives rise to $\delta(\kappa+\kappa')$ type correlations when $\epsilon\rightarrow 0$ (whereas the first line corresponds to $\delta(\kappa-\kappa')$) hence we restrict our attention to the last line of \eqref{ThetaCorrel1}. To proceed, we approximate the box-function originating from the Heaviside Theta function (c.f. \eqref{HeavisideThetas}) by a continuous, analytic and exponentially decaying function, for instance the difference of two hyperbolic tangents $\tilde{\Theta}_\text{H}^{\epsilon,\delta}(x)=\tanh((\epsilon/2-x)/\delta)-\tanh((-\epsilon/2-x)/\delta)$, which allows us to consider the entire $k-p$ plane as the range of integration as well as using the stationary phase approximation to evaluate the oscillatory integrals. We also introduce the scale $\ell$, and for the sake of transparency consider the double summation wrt. the rapidity variables $\kappa_n$ and $\kappa_{n'}$ and use a test function $\mathcal{F}(\kappa_n,\kappa_{n'})$, which can be arbitrary or be thought as a peaked one at specific $\kappa$ and $\kappa'$ values. That is, we eventually write that
\beq
\begin{split}
\sum_{n,n'}\phantom{}_\text{FF}\langle\Psi|q_{\kappa_n}(\ell x,\ell t)q_{\kappa_{n'}}(\ell x',\ell t)|\Psi\rangle^{c*}_\text{FF}\,\,\mathcal{F}(\kappa_n,\kappa_{n'})
&=\sum_{n,n'}\int \frac{\dd k}{2\pi}\frac{\dd p}{2\pi}\mathcal{K}_2(k,p)e^{-\ri \ell(x-x')(k-p)}e^{2\ri \ell t(k^2-p^2)}\times\\
&\quad\times\tilde{\Theta}_\text{H}^{\epsilon, \delta}\left(\frac{k+p-2\kappa_n)}{2}\right)\tilde{\Theta}_\text{H}^{\epsilon, \delta}\left(\frac{k+p+2\kappa_{n'})}{2}\right)\,\mathcal{F}(\kappa_n,\kappa_{n'})\\
&=\frac{1}{4\pi\ell t}\sum_{n,n'}\mathcal{K}_2(k^*,p^*)\mathcal{F}(\kappa_n,\kappa_{n'})\times\\
&\quad\tilde{\Theta}_\text{H}^{\epsilon, \delta}\left(\frac{k^*+p^*-2\kappa_n)}{2}\right)\tilde{\Theta}_\text{H}^{\epsilon, \delta}\left(\frac{k^*+p^*+2\kappa_{n'})}{2}\right)\,,
\end{split}
\eeq
where $\langle.\rangle^{c*}$ means connected correlations originating from $\kappa, -\kappa$ modes, $k^*=p^*=(x-x')/(4t)$ from the saddle-point and it is easy to see that  no exponential factors remain after the stationary-phase approximation. 
Denoting $(x-x')/(4t)$ by $\zeta$ and simplifying out notation by $\mathcal{K}_2(\kappa,\kappa)=\mathcal{K}_2(\kappa)$ and $\mathcal{F}(\kappa,\kappa)=\mathcal{F}(\kappa)$, we now perform the $\delta\rightarrow0$ limit in the approximating box function and thus end up with the usual Heaviside Theta functions allowing us to write 
\beq
\begin{split}
\sum_{n,n'}\phantom{}_\text{FF}\langle\Psi|q_{\kappa_n}(\ell x,\ell t)q_{\kappa_{n'}}(\ell x',\ell t)|\Psi\rangle^{c*}_\text{FF}\,\,\mathcal{F}(\kappa_n,\kappa_{n'})
&=\sum_{n,n'}\frac{\mathcal{K}(\zeta)}{4\pi\ell t} \Theta_\text{H}(\epsilon/2-|\zeta-\kappa_n|)\Theta_\text{H}(\epsilon/2-|\zeta+\kappa_{n'}|)\,\mathcal{F}(\zeta)\\
&=\epsilon^2\!\sum_{n,n'}\frac{\mathcal{K}(\zeta)}{4\pi\ell t} \frac{\Theta_\text{H}(\epsilon/2-(\zeta-\kappa_n))-\Theta_\text{H}(-\epsilon/2-(\zeta-\kappa_n))}{\epsilon}\\
&\qquad\times\frac{\Theta_\text{H}(\epsilon/2-(\zeta+\kappa_{n'}))-\Theta_\text{H}(-\epsilon/2-(\zeta+\kappa_{n'}))}{\epsilon}\mathcal{F}(\zeta)\,,
\end{split}
\eeq
which in the $\epsilon\rightarrow 0$ limit becomes
\beq
\int \dd \kappa\, \dd \kappa' \frac{1}{4\pi\ell t}\mathcal{K}_2(\zeta)\delta(\zeta-\kappa)\delta(\zeta+\kappa')\mathcal{F}(\zeta)=\int \dd \kappa\, \dd \kappa' \frac{1}{4\pi\ell t}\mathcal{K}_2(\kappa)\delta(\frac{x-x'}{4t}-\kappa)\delta(\kappa+\kappa')\mathcal{F}(\kappa)\,,
\eeq
where in the last line we recalled that $\zeta=(x-x')/(4t)$. Using that $t>0$ and $\delta(\kappa-z/t)/t=\delta(\kappa t-z)$ we have that in the $t\rightarrow 0$ limit
\beq
\int\!\! \dd \kappa\, \dd \kappa'\phantom{}_\text{FF}\langle\Psi|q_{\kappa_n}(\ell x,\ell t)q_{\kappa_{n'}}(\ell x',\ell t)|\Psi\rangle^c_\text{FF}\,\mathcal{F}(\kappa)=\frac{1}{\pi \ell}\int \!\!\dd \kappa\, \dd \kappa' \!\left(\mathcal{K}_2(\kappa)\delta(\kappa+\kappa')\delta(x-x')+\mathcal{K}_1(\kappa)\delta(x-x') \delta(\kappa-\kappa')\right)\mathcal{F}(\kappa)\\
\eeq
where  the $\delta(\kappa-\kappa')$ type correlations originating from the 1st line of the rhs. in \eqref{ThetaCorrel1} are not relevant for our purposes and exploiting the arbitrariness of the test function $\mathcal{F}(\kappa)$, we  end up with
\beq
\lim_{t \rightarrow 0}\lim_{\epsilon\rightarrow 0}\lim_{\ell \rightarrow \infty}\ell\times\phantom{}_\text{FF}\langle\Psi|q_{\kappa_n}(\ell x,\ell t)q_{\kappa_{n'}}(\ell x',\ell t)|\Psi\rangle^{c*}_\text{FF}= \frac{1}{\pi}\mathcal{K}_2(\kappa)\delta(\kappa+\kappa')\delta(x-x')\,.
\label{InitialFFCorrelationThetaTheta}
\eeq
It is also easy to show that the limits $ \lim_{t \rightarrow 0}\lim_{\epsilon\rightarrow 0}$ can be exchanged as well. 
\end{widetext}

\subsubsection{Estimating long-range correlations in the LL model}
As explained, curved hydrodynamic trajectories and the particular correlations in the initial state can give rise to long range correlations for current multi-point functions after the bi-partite state and modify the time dependence of the scaled cumulants of the conserved charge on the original quench problem. We now use a simply and intuitive picture in the spirit of BMFT to provide a prediction for the order at which deviations attributed to this effect appear. We stress again that the origin of this phenomenon is twofold: fluctuating hydrodynamic modes have non-trivial initial correlations due to the pair structure of the initial state as modes with opposite rapidity are correlated. Second, although these modes have opposite velocities at short times, due to interactions, there exists a finite region in rapidity space $[–\Lambda, \Lambda]$ in which every normal mode will eventually have the same sign of its effective velocity in the NESS, therefore initially correlated modes build up correlations between observable placed at the same spatial point but at different times. As confirmed by Fig. \ref{fig:Big} a), the effective velocity in this region can simply be approximated as 
\beq
v^\text{eff}(k, \xi=0,\lambda)\approx b_1*(k- \Lambda) \quad \text{if } k\in [-\Lambda, \Lambda]
\label{vEffApprox}
\eeq
where $\Lambda\propto b_2\lambda $ for small $\lambda$ (c.f. Fig. \ref{fig:Big} c)) and we consider the $e^{\lambda Q|_{0}^\infty}|\Psi\rangle$ state. Additionally $b_1$ is an $\mathcal{O}(1)$ quantity mildly depending on $\lambda$ and $b_2$ is an $\mathcal{O}(1)$ factor, and both $b_1$ and $b_2$ depend on the LL interaction strength $c$ and the density $d$, as supported by extensive numerical studies.

In the following let us estimate the contribution of such correlated normal modes first for the current-current 2pt function $\langle\langle  j(0,t)j(0,t')\rangle\rangle^{c*}$ at a finite value of the counting field $\lambda$, where the expectation value taken at the Euler scale and is understood in the BMFT sense. That is, 
correlations spread according to Euler hydrodynamics characterized by the initial density matrix $\rho_\text{GGE}^\text{(L)}\otimes \rho_\text{GGE}^\text{(R)}e^{2\lambda Q}$ where the superscript (L/R) means that the density matrix is non-trivial only on the left/right side of the system, but the initial correlations are an external output, which we obtained in the previous part of this section (VI.B.2).

According to BMFT we regard the currents in $\langle\langle  j(0,t)j(0,t')\rangle\rangle^{c*}$ as fluctuating fields, which are functionals of conserved densities. These densities themselves are fluctuating fields, with arguments the macroscopic space and time coordinates, and hence we write
\beq
\langle  j(0,\ell t)j(0, \ell t')\rangle^{c*}=\langle\langle  j[\vec q(0, t)]j[\vec q(0, t')]\rangle\rangle^{c*}\,.
\eeq

We first recall that the current and charge expectation values in GGEs can be expressed as
\beq
q( x, t)=\int \dd \kappa\, \left(p'(\kappa)\right)^\text{dr} \theta(\kappa, \xi,\lambda)
\eeq
hence for the current we can write
\beq
\begin{split}
&j[\vec q( x, t)]=\\
&\int \!\!\dd \kappa\, \left(E'(\kappa)\right)^\text{dr} \!\theta(\kappa, \xi,\lambda)=\!\!\int \!\! \dd \kappa\, E'(\kappa) \theta(\kappa, \xi,\lambda) 1^\text{dr}(\kappa, \xi,\lambda)\\
\end{split}
\eeq
where for the the LL model $p'=1$, hence $\left(p'(\kappa)\right)^\text{dr}=1^\text{dr}$. Therefore, we can simply write
\beq
\langle  j(0,\ell t)j(0,\ell t')\rangle^{c*}\!\!= \!\!\int\!\! \dd \kappa\,\dd \kappa' E'(\kappa)E'(\kappa')\langle  q_{\kappa}(0,  t)q_{\kappa'}(0,  t')\rangle^{c*}
\label{jj-qq-strong}
\eeq
(which is specific to the particle density and current) and consequently also
\beq
\begin{split}
&\langle\langle   j[\vec q(0, t)]j[\vec q(0, t')]\rangle\rangle^{c*}=\\
&\quad=\int \! \dd \kappa\,\dd \kappa' E'(\kappa)E'(\kappa')\langle \langle  q_{\kappa}(0,  t)q_{\kappa'}(0,  t')\rangle\rangle^{c*}\,,
\label{jj-qq}
\end{split}
\eeq
where we used the rapidity-resolved particle density.

In the spirit of BMFT, correlations are transported ballistically. This means that the time evolution of the fluctuating fields is dictated by Euler hydrodynamics or, equivalently, the hydrodynamic normals modes $\theta(\kappa)$ evolve according to the Euler characteristics. However, we may write $q_\kappa(x,t)\approx \theta (\kappa, x,t)$ due to the following reason.
As supported by extensive numerical studies (cf. Appendix \ref{App:NumericallyInvestigatedCurvedTraj} and Table \ref{table1}), the quantity 
\beq
1^\text{dr}(\kappa, \xi, \lambda)\approx b_3
\label{1drApprox}
\eeq
i.e., a constant (depending on $c$ and $d$), to a very good approximation: for rapidities $\kappa \in [-\Lambda,\Lambda]$, $b_3$ is typically 2-5 for parameters $c$ and $d$ and $\lambda$ with generic $\mathcal{O}(1)$ values. Importantly, by varying the ray parameter $\xi$ between $-\infty$ and $0$, or $0$ and $\infty$, $b_3$ changes approximately 20-25\% if $\lambda\propto \mathcal{O}(1)$ (cf. Fig. \ref{fig:Big} b) and only a few percent if $\lambda\propto \mathcal{O}(10^{-1})$. In other words, $1^\text{dr}$ can regarded as a constant allowing for interchanging $q$ and $\theta$ in \eqref{jj-qq} and hence writing 
\beq
\begin{split}
&\langle\langle  j[\vec q(0, t)]j[\vec q(0, t')]\rangle\rangle^{c*}\approx \\
&\approx\!\int\!\! \dd \kappa\,\dd \kappa' E'(\kappa)E'(\kappa')\langle\langle  q_{\kappa}(X(0,\kappa),0)q_{\kappa'}(X(0,\kappa'),0)\rangle \rangle ^{c*},
\label{q-theta-approx}
\end{split}
\eeq
where $q_k$ evolve according to the characteristic equation and thus $X(\tau, \kappa)$ is the Euler characteristics, i.e., solutions of the equation $\frac{\text{d}X}{\text{d}t}=v^\text{eff}(\kappa, X/t,\lambda)$,  with properties $X(\tau= t, \kappa)=0$. 
We now exploit the fact that the Euler characteristics or the  trajectories of small velocity hydrodynamic modes dominantly fall into a space-time region that corresponds to the NESS with $\xi=0$ as indicated by Fig \ref{fig:CurvedTrajectories} and Fig. \ref{fig:Big} d). 
This way, we have that $X(\tau, \kappa)\approx v^\text{eff}(\kappa,\xi=0,\lambda)( \tau- t)$ and hence we can eventually write that
\beq
\begin{split}
&\langle\langle  j[\vec q(0, t)]j[\vec q(0, t')]\rangle\rangle^{c*}\approx \int \dd \kappa\,\dd \kappa' E'(\kappa)E'(\kappa')\times\\
&\langle\langle  q_{\kappa}(- v^\text{eff}(\kappa,\xi=0,\lambda)t,0)q_{\kappa'}(- v^\text{eff}(\kappa',\xi=0,\lambda)t',0)\rangle^{c*}\,.
\label{q-theta-straightline-approx}
\end{split}
\eeq

Approximating $v^\text{eff}$ according to \eqref{vEffApprox},
we can evaluate the expectation value in the initial state, more precisely, at short times on the Euler scale. Using also the approximate expression for such initial density correlations \eqref{InitialFFCorrelationThetaTheta}, we can proceed as
\begin{widetext}
\beq
\begin{split}
\langle\langle  j[\vec q(0, t)]j[\vec q(0, t')]\rangle\rangle^{c*}&\approx\frac{1}{\ell}\int \dd \kappa\,\dd \kappa' E'(\kappa)E'(\kappa')\delta(\kappa+\kappa')\delta(v^\text{eff}(\kappa,\xi=0,\lambda)t-v^\text{eff}(\kappa',\xi=0,\lambda)t')\\
&\qquad\qquad\times\left(\Theta_\text{H}(-v^\text{eff}(\kappa,\xi=0,\lambda)t)\mathcal{K}_2^\text{a}(\kappa)+\Theta_\text{H}(v^\text{eff}(\kappa,\xi=0,\lambda)t)\mathcal{K}_2^\text{b}(\kappa)\right)\\
&=\frac{1}{\ell}\int \dd \kappa\, E'(\kappa)E'(-\kappa)\delta(v^\text{eff}(\kappa,\xi=0,\lambda)t-v^\text{eff}(-\kappa,\xi=0,\lambda)t')\\
&\qquad\qquad\times\left(\Theta_\text{H}(-v^\text{eff}(\kappa,\xi=0,\lambda)t)\mathcal{K}_2^\text{a}(\kappa)+\Theta_\text{H}(v^\text{eff}(\kappa,\xi=0,\lambda)t)\mathcal{K}_2^\text{b}(\kappa)\right)\\
\end{split}
\eeq
which further simplifies to
\beq
\begin{split}
\langle\langle  j[\vec q(0, t)]j[\vec q(0, t')]\rangle\rangle^{c*}&\approx\frac{1}{\ell}\int \dd \kappa\, E'(\kappa)E'(-\kappa)\frac{\delta\left(\pm\frac{t'-t}{t+t'}\frac{\Lambda}{b_1}-\kappa\right)}{|\frac{\dd}{\dd \kappa}v^\text{eff}(\kappa,\xi=0,\lambda)t-\frac{\dd}{\dd \kappa}v^\text{eff}(-\kappa,\xi=0,\lambda)t'|}\\
&\qquad\qquad\times\left(\Theta_\text{H}(-v^\text{eff}(\kappa,\xi=0,\lambda)t)\mathcal{K}_2^\text{a}(\kappa)+\Theta_\text{H}(v^\text{eff}(\kappa,\xi=0,\lambda)t)\mathcal{K}_2^\text{b}(\kappa)\right)\\
&=-\frac{4}{\ell} \left(\frac{t'-t}{t+t'}\frac{\Lambda}{b_1}\right)^2 \frac{1}{b_1|t-t'|}\mathcal{K}_2^{(2\lambda)}\left(\pm\frac{t'-t}{t+t'}\frac{\Lambda}{b_1}\right)\,,
\end{split}
\eeq
\end{widetext}
due to \eqref{vEffApprox} 
where a and b mean either $\emptyset,2\lambda$ or $2\lambda,\emptyset$ and $\mathcal{K}_2^\text{a/b}(x)=\mathcal{K}_2^\text{a/b}(x,x)$ defined by \eqref{calK} via $K(k)$ or $K(k)e^{2\lambda}$.  At this stage some properties of the unknown function $\mathcal{K}_2^{2\lambda}$ are relevant. For free fermion quenches $\mathcal{K}_2^{2\lambda}(0)=0$, i.e., at small rapidities the function cannot be a constant, and the mildest possible behavior is the quadratic one, which is a consequence of the fact that fermionic $K$-functions are odd in accordance with fermionic statistics. In fact, odd $K$-functions with a singularity at $\kappa=0$ can also define integrable quenches c.f. the Tonks-Girardeau limit of the LL BEC quench \cite{Kormos_2014} with $K(\kappa)=2d/\kappa$ and hence $\mathcal{K}_2^{2\lambda}(0)=0$ with a zero first derivative also in this case. Additionally, we also note that for free fermions $\mathcal{K}_2^{2\lambda}(\kappa)=\theta^{2\lambda}(\kappa)\left(1-\theta^{2\lambda}(\kappa)\right)$. 

It is plausible to assume that $\mathcal{K}^{2\lambda}(\kappa)$ is an even function with quadratic behavior at the origin for the LL case. Namely
one can regard the Tonks-Girardeau limit of the LL model and the effective fermionic statistic of the excitations in the LL model at generic $c$. Additionally, the resolution of the dynamical structure factor, in terms of a single rapidity-integral, after the BEC quench, involves the factor $\theta(\kappa)\left(1-\theta(\kappa)\right) \rho_t(\kappa)^3$ \cite{De_Nardis_2016} which has a quadratic behavior at small $\kappa$.
Nevertheless, to keep the following discussion general we shall write $\mathcal{K}_2^{2\lambda}(\kappa)\approx \mathcal{K}^{2\lambda}(0)+b_4(\lambda) \kappa^2$, which have a finite value at $\lambda=0$ (i.e., from $\mathcal{K}(\kappa)\approx \mathcal{K}_2(0)+b_4(0)\kappa^2$), and which means that the leading order contribution to the current-current correlations at the Euler-scale is given by 
\begin{widetext}
\beq
\begin{split}
\langle  j(0,\ell t)j(0, \ell t')\rangle^{c*}&\approx -\frac{4}{\ell} \left(\frac{1}{t+t'}\frac{\Lambda}{b_1}\right)^2 \frac{|t-t'|}{b_1}\mathcal{K}^{2\lambda}(0)-\frac{2 b_4}{\ell} \left(\frac{1}{t+t'}\frac{\Lambda}{b_1}\right)^4 \frac{|t-t'|^3}{b_1}\\
&=-\frac{1}{\ell}\frac{|t'-t|}{(t+t')^2} \lambda^2 b_1' -\frac{1}{\ell}\frac{|t'-t|^3}{(t+t')^4} \lambda^4 b_2'\,,
\label{CF}
\end{split}
\eeq
where $b_1'$ and $b_2'$ are $\mathcal{O}(1)$ numbers independent on $\lambda$ (keeping in mind that $b_1'$ is plausibly zero) and we have exploited that fact that $\Lambda\approx \lambda b_2$ for small $\lambda$-s. 
We can also estimate the correction to the 2nd scaled cumulant of the charge at finite counting field by integrating the current 2pt function wrt. time. This is done by taking into account the Euler-scaling as well, and the fact the contributions on the left and right the vertical contours are equal:
\beq
\begin{split}
\Delta c^{\lambda}_2&=2\lim_{\ell \rightarrow \infty}\frac{1}{\ell T}\int_0^{\ell T}\int_0^{\ell T} \dd t\,\dd t' \langle  j(0,t)j(0,t')\rangle^{c*}=2\lim_{\ell \rightarrow \infty}\frac{\ell}{T}\int_0^{T}\int_0^{T} \dd t\,\dd t' \langle  j(0,\ell t)j(0,\ell t')\rangle^{c*}\\
&\approx 2\lim_{\ell \rightarrow \infty}\frac{\ell}{ T}\int_0^{T}\int_0^{T} \dd t\,\dd t'\frac{1}{\ell}
\frac{|t'-t|}{(t+t')^2} \lambda^2 b_1'+2\lim_{\ell \rightarrow \infty}\frac{\ell}{ T}\int_0^{T}\int_0^{T} \dd t\,\dd t'\frac{1}{\ell}
\frac{|t'-t|^3}{(t+t')^4} \lambda^4 b_2'\\
&=-4(1-\ln2) b_1' \lambda^2-4(5/6-\ln2) b_2' \lambda^4\,,
\end{split}
\eeq
where it is important to stress that the temporal scaling of the corrections is regular (i.e., it is time-independent) and the fact that it is plausible that $b_1'=0$.
Given the expansion of the SCGF we can write
\beq
c^{\lambda}_2+\Delta c^{\lambda}_2=\frac{\partial^2}{\partial \lambda^2} 2\tilde{f}_\text{dyn}(\lambda)=c_2+c_3 \lambda +\frac{c_4 \lambda ^2}{2}+\frac{c_5 \lambda ^3}{6}+\frac{c_6 \lambda ^4}{24}+..
\label{last}
\eeq
\end{widetext}
where $\tilde{f}_\text{dyn}(\lambda)$ denotes the "correct" dynamical part of the SCGF that includes also the corrections due to long-range correlations, i.e., $\tilde{f}_\text{dyn}(\lambda)=f_\text{dyn}(\lambda)+...\,$. From \eqref{last}
we can deduce the correction $\Delta c^{\lambda}_2$ eventually modifies 
$c_6$ (or possibly $c_4$). Since at finite $\lambda$ also $ c^{\lambda}_2$ depends on the higher scaled cumulants, the identification of the precises correction to such higher cumulants  is non-trivial and  we leave this task for later works also aiming the more precise computation of corrections to higher point correlation functions and further higher cumulants. We conclude by stressing that long-range correlations modify the SCGF either at the 4th or the 6th order, meaning that the scaled cumulants $c_n$ 
$n\geq 6$ (or $n\geq4$) obtain correction wrt. the predictions of the BFT flow equations \eqref{Flow1} and \eqref{Flow2}.

\section{Conclusion} We have derived, from general principles of many-body physics and integrability, a formula for the fluctuation of the number of particles in a gas on a large interval of size $X$, a large time $T$ after an integrable quench, in the ballistic scaling limit, with $X\gg T$. This formula reproduces a recent conjecture \cite{Bertini_2023_tFCS, Bertini_2024_tFCS}. We have shown how it is the physics of hydrodynamic long-range correlations that largely determine the fluctuations: the scaling with $X$ accounts for a modified thermodynamic function (evaluated exactly in integrable systems) due to such correlations produced by the quench, and its non-trivial dependence on $T$ is solely a consequence of long-range correlations at different times. Crucially, our derivation shows how this genuinely far from equilibrium dynamical effect is recast into current fluctuations at a point $x=0$ on a time interval $[0,T]$ in a special NESS. The dynamics may be modified by further long-range ``waft effects" (neglected up to now in the literature) where correlated modes are transported to $x=0$ by the flow of the NESS. We have argued that in integrable systems with fundamental excitations of fermionic statistics, waft effects may only affect cumulants of order 6 or higher, in a way that cannot be currently evaluated. We expect that waft effects do not alter cumulants in full generality when the model admits at most two hydrodynamic velocities; and in integrable systems when hydrodynamic velocities are bounded from below in magnitude (for instance in hard rods systems).
Many of the arguments we have made are in fact immediately applicable beyond integrability, see \cite{LongPaper}. It would be interesting to extend our arguments to general $\alpha = X/T$ and other geometries, and to other charges or to the entanglement entropies; and to evaluate the proposed long-range effects modifying the dynamics of higher cumulants using BMFT. Extensive numerical checks of the formula would be desired as well, especially in the context of non-integrable models.

\begin{acknowledgments}

We thank discussions with Bruno Bertini, Colin Rylands, Pasquale Calabrese, Filiberto Ares, Shachar Fraenkel and Lenart Zadnik. The work of BD and DXH has been supported by the Engineering and Physical Sciences Research Council (EPSRC) under grant numbers EP/W010194/1 and EP/Z534304/1 (UK Research and Innovation Horizon Europe Guarantee, Advanced Grant Scheme).
\end{acknowledgments}

\bibliography{FCSBFTPRRRegularArticleAcceptedArXiv.bib}

\appendix

\section{Analysis of a commutator}
\label{App:Commutator}

Consider, as in the main text (with $X\to\infty$),
\begin{equation}
    Q|_{0}^\infty = \int_0^\infty\dd x\,q(x,0),\quad J_0|_0^T = \int_0^T\dd t\,j(0,t).
\end{equation}
In order to underpin the arguments in the main text, we would like to show that $[Q|_0^\infty,J_0|_0^T]$ is supported around the origin $x=0$, with a norm
\begin{equation}
    ||\,[Q|_0^\infty,J_0|_0^T]\,|| = o(T).
\end{equation}
We are not aware of general results allowing us to show this. However, a possible line of arguments is as follows.

We propose that $j(0,t)$ is composed of linear combinations of ``local enough" operators supported on $x\in[-vt,vt]$ (for appropriate $v$), which become ``thinner" as $t$ increases, in the sense that they become closer to the identity operator (see \cite{Ampelogiannis2023}). In order to make this slightly more accurate, consider the projectors $\mathbb P_{x>0}$, $\mathbb P_{x<0}$ on operators supported on the right and left half of space (see e.g.~\cite{Ampelogiannis2023} for definitions of such projectors). We then propose two loosely-stated principles concerning the operator $(1-\mathbb P_{x>0} - \mathbb P_{x<0})j(0,t)$: (i) it is supported around the origin $x=0$, and (ii) it becomes closer to the identity as $t$ increases. As a consequence,
\begin{equation}
    B(t) = [Q|_0^\infty,(1-\mathbb P_{x>0} - \mathbb P_{x<0})j(0,t)]
\end{equation}
is supported around the origin, and $\int_0^T\dd t\, B(t) = o(T)$ (in an appropriate norm). This is sufficient to obtain the desired result:
\begin{eqnarray*}
    [Q|_0^\infty,j(0,t)] &=& B(t) + [Q|_0^\infty,\mathbb P_{x>0}j(0,t)]\\
    & &\qquad\,\,+ [Q|_0^\infty,\mathbb P_{x<0}j(0,t)]\\
    &=& B(t) + [Q|_0^\infty,\mathbb P_{x>0}j(0,t)] \\
    &=& B(t) + [Q,\mathbb P_{x>0}j(0,t)] \\
    &=& B(t) + \mathbb P_{x>0}[Q,j(0,t)]\\
    &=& B(t)
\end{eqnarray*}
where we used the fact that $Q$ is ultra-local thus commute with the projection, and that the current $j$ is invariant under the action of $Q$ (preserves the particle number), $[Q,j(0,t)]=0$.

\section{Implications of the hydrodynamic fluctuation relations}\label{FluctuationRelations}
In this subsection we demonstrate that the hydrodynamic fluctuation relations \cite{Esposito_2009, Bernard_2016,Perfetto_2020} which were derived for thermal states strongly confirms the validity of performing steps such as $e^{A+B}\approx e^{A}e^{B}$ for the eventual quench problem in the LL model and consequently in a broad class of interacting integrable systems as well. More precisely, we show that fluctuation relations are compatible with writing Eq.~(10) in the main text in thermal or GGE ensembles indirectly justifying the legitimacy of such steps in GGEs from which we infer the validity of Eq.~(10) in the main text for the quench problem as well. 

The main idea is to consider the FCS in a GGE (rather than the quench) and exploit its time-independence. 
In particular using the contour deformation from the main text we may write $F(\lambda, X,T)=T f_\text{dyn}^\text{L}(\lambda)+T f_\text{dyn}^\text{R}(\lambda)+X f^0(\lambda)$ with $X \gg T \gg 1$, up to sub-leading corrections as in the main text, with
\beq
\begin{split}
 f_\text{dyn}^\text{L}(\lambda)=&\lim_{T\rightarrow \infty}T^{-1}\lim_{X\rightarrow \infty}\ln\frac{\Tr[e^{\lambda J_0|_0^T+\lambda Q|_0^{X/2}(0)} \rho_\text{GGE}]}{\Tr[e^{\lambda Q|_0^{X/2}(0)}\rho_\text{GGE}]}\\
 f_\text{dyn}^\text{R}(\lambda)=&\lim_{T\rightarrow \infty}T^{-1}\lim_{X\rightarrow \infty}\ln\frac{\Tr[e^{\lambda Q|_{X/2}^{X}(0)-\lambda J_X|_0^T} \rho_\text{GGE}]}{\Tr[e^{\lambda Q|_{X/2}^{X}(0)} \rho_\text{GGE}]}\,,
\label{FirstSplitAppendix}
\end{split}
\eeq
where we may recall the cluster property of GGEs. $f^0(\lambda)$ is the SCGF in a GGE, which must be equal to $\lim_{X\rightarrow\infty}F(\lambda,T,X)/X$ due to the stationary property of GGEs implying that 
\beq
\begin{split}
0=&\lim_{X\rightarrow \infty}\ln\frac{\Tr[e^{\lambda J_0|_0^T+\lambda Q|_0^{X/2}(0)} \rho_\text{GGE}]}{\Tr[e^{\lambda Q|_0^{X/2}(0)}\rho_\text{GGE}]}\\
&+\lim_{X\rightarrow \infty}\ln\frac{\Tr[e^{\lambda Q|_{X/2}^{X}(0)-\lambda J_X|_0^T} \rho_\text{GGE}]}{\Tr[e^{\lambda Q|_{X/2}^{X}(0)} \rho_\text{GGE}]}
\label{ugly}
\end{split}
\eeq
must hold.
Rewriting $f_\text{dyn}^\text{L}(\lambda)$ as
\beq
 f_\text{dyn}^\text{L}(\lambda)=\lim_{T\rightarrow \infty}T^{-1}\ln\frac{\Tr[e^{\lambda J_0|_0^T}e^{\lambda Q|_0^{\infty}(0)} \rho_\text{GGE}]}{\Tr[e^{\lambda Q|_0^{\infty}(0)}\rho_\text{GGE}]}\,,
\eeq
it also follows that 
\beq
f_\text{dyn}^\text{L}(\lambda)+f_\text{dyn}^\text{R}(\lambda)=0\,.
\eeq
The important recognition is that, by definition, the quantity $f_\text{dyn}^\text{L}(\lambda)$ equals the SCGF of the $i$th current, in a NESS after a bi-partite quench where the left semi-infinite system is characterized by a GGE, $\rho_\text{GGE}$; and the right half is by $e^{\lambda Q}\rho_\text{GGE}$. Importantly, unlike in the quench problem, $e^{\lambda Q}\rho_\text{GGE}$, is a biased GGE where $\beta^*$ becomes shifted to $\beta^*-\lambda$ (and not $2\lambda$), by definition. We have assumed that the system locally equilibrates to the NESS of this bi-partitioning protocol as well as the lack of strong temporal long-range correlations in NESS, which is legitimate since in GGEs quasi-particles with opposite momentum are not correlated. In other words, we have found that 
\beq
f_\text{dyn}^\text{L}(\lambda)=f^\text{bpp}(\lambda)\,,
\eeq
where $f^\text{bpp}(\lambda)$ is the SCGF if the $i$th current in the NESS of the GGE emerging after the aforementioned partitioning protocol and which has been studied in the literature \cite{Esposito_2009, Bernard_2016}.
In particular $f^\text{bpp}(\lambda)$ satisfies  the hydrodynamic fluctuation relations, i.e.,
\beq
f^\text{bpp}(\lambda)=f^\text{bpp}(\beta^\text{R}-\beta^\text{L}-\lambda)
\eeq
and since in our case $\beta^\text{L}=\beta$ and $\beta^\text{R}=\beta-\lambda$, we find that 
\beq
f_\text{dyn}^\text{L}(\lambda)=f^\text{bpp}(0)=0
\eeq
guaranteeing \eqref{ugly},
at least for large $T$ (the same argument can be made for $f_\text{dyn}^\text{R}$).
In other words we have shown that the hydrodynamic fluctuation relations imply the expected invariance of the charge FCS in GGEs if the step Eq.~(10) is performed and no strong temporal correlations affecting current fluctuations are assumed. That is, the fulfilment of these assumptions in GGEs is consistent with physical requirements, such as the implication of hydrodynamic fluctuation relations or the invariance of GGEs. These findings make performing Eq.~(10) for the quench problem very plausible.

\section{Emergence of the biased GGE via Quench Action}
\label{QAComputation}
In the following we show that after the biased integrable quench the steady-state value of local operators is described by a biased GGE in any interacting integrable model, that is
\beq
\begin{split}
&\lim_{t\rightarrow \infty}\frac{\langle \Psi|e^{\lambda/2\, Q}O(0,t)e^{\lambda/2\, Q}|\Psi\rangle}{\langle \Psi|e^{\lambda\, Q}|\Psi\rangle}=\\
&=
\lim_{t\rightarrow \infty}\frac{\langle \Psi|O(0,t)e^{\lambda Q}|\Psi\rangle}{\langle \Psi|e^{\lambda Q}|\Psi\rangle}=\frac{1}{Z^{2\lambda}}\Tr \left[O(0,0)e^{2\lambda Q}\rho_\text{GGE}\right]
\label{2BetaGGE}
\end{split}
\eeq
where $\rho_\text{GGE}$ is the GGE that specifies that steady-state, or on other words, where the expectation value of any local operator $O$ converges,
\beq
\lim_{t\rightarrow \infty}\frac{\langle \Psi|O(0,t)|\Psi\rangle}{\langle \Psi|\Psi\rangle}=\frac{1}{Z}\Tr \left[O(0,0)\rho_\text{GGE}\right]\,.
\eeq
To show Eq. \eqref{2BetaGGE} we shall use the reasonings of the Quench Action (QA) method \cite{caux2013time, caux2016quench}; in fact, the derivation generally applies to arbitrary integrable models as long as the fluctuations of the integrated conserved density $Q|_{-\ell/2}^{\ell/2}$ are extensive wrt. $\ell$. For simplicity, however, let us assume only one particle species.
Writing the expectation value using the eigenstates of the post-quench Hamiltonian we obtain
\begin{widetext}
\begin{equation}
\frac{\langle \Psi|O(0,t)e^{\lambda Q}|\Psi\rangle}{\langle \Psi|e^{\lambda Q}|\Psi\rangle}=\frac{1}{\langle\Psi|e^{\lambda Q}|\Psi\rangle}\sum_{\Phi,\Phi'}e^{-\epsilon_{\Phi}^{*}-\epsilon_{\Phi'}}e^{\ri(\omega_{\Psi}-\omega_{\Psi'})t}\langle\Phi|O(0,0)e^{\lambda Q}|\Phi'\rangle\,,
\label{DE-FCSDerivation1}
\end{equation}
where $\epsilon_\Phi=-\log{\langle\Phi |\Psi\rangle}$ are the logarithmic overlaps. Following the logic of the QA method, we can replace one summation by a functional integral over root distributions assuming that the size of the entire system $L$ is very large and is eventually sent to infinity. This way we may write
\begin{eqnarray}
\frac{\langle \Psi|O(0,t)e^{\lambda Q}|\Psi\rangle}{\langle \Psi|e^{\lambda Q}|\Psi\rangle}=\sum_{\Phi}\!\int\!\mathcal{D}[\rho]e^{S_{YY}[\rho]} \!\left[  e^{-\epsilon_{\Phi}^{*}-\epsilon[\rho]} e^{\ri(\omega_{\Psi}-\omega[\rho])t}  \langle\Phi|  O(0,0)e^{\lambda Q}|\rho\rangle +\Phi\leftrightarrow\rho\vphantom{\frac{OOOOO}{OOOOO}}\right]\times \frac{1}{\langle\Psi |e^{\lambda Q}|\Psi \rangle},\quad\quad\quad
\label{DE-FCSDerivation2}
\end{eqnarray}
\end{widetext}
where $S_{YY}$ is the Yang-Yang entropy of the root distribution whose exponential gives the number of microstates with the same root distribution. Note that this quantity is proportional to the system size $L$. 
The usual reasoning of QA comes from the recognition that when $t$ is sent to $\infty$, because of the oscillatory factor and the behaviour of matrix elements in the thermodynamic limit, the summation over the eigenstates $\Phi$ can be terminated and replaced by a single eigenstate that corresponds to the macro-state $\rho$, i.e., 
\begin{equation}
\begin{split}
&\frac{\langle \Psi|O(0,\infty)e^{\lambda Q}|\Psi\rangle}{\langle \Psi|e^{\lambda Q}|\Psi\rangle}=\\
&\quad\quad=\frac{\int\mathcal{D}[\rho]O[\rho]e^{S_{YY}[\rho]-2 \text{Re}\,\epsilon[\rho]+L\int \text{d}k\, \lambda q(k) \rho(k) }}{\int\mathcal{D}[\rho]e^{S_{YY}[\rho]-2 \text{Re}\,\epsilon[\rho] +L\int \text{d}k\, \lambda q(k) \rho(k)}}\,,
\label{DE-FCSDerivation4}
\end{split}
\end{equation}
where we also exploited the fact the $Q$ is an extensive conserved quantity and hence can be naturally rewritten in terms of the root-distributions. In the thermodynamic limit, the functional integral over the root densities can be replaced by its saddle-point value, since not only the Yang-Yang entropy, but also the conserved charge and by definition, the extensive part of the logarithmic overlap $\epsilon[\rho]$ scale with the system size $L$. The main ingredient which we further  utilize is that a local operator cannot shift the saddle point, which, accordingly, is determined by the denominator. In the usual fashion of QA, we can write the exponential in denominator as
\beq
-L\int_0^{\infty} \text{d}k\, g(k)\rho(k)+s_{YY}[\rho(k)]+L\int_{-\infty}^{\infty} \text{d}k\, \lambda q(k) \rho(k)\,,
\eeq
where $s_{YY}$ is the Yang-Yang entropy density and for integrable quenches, 
\beq
-2\text{Re}\,\epsilon[\rho]=L\int_0^{\infty} \dd k \, g(k)\rho(k)\,.
\eeq
Note that the lower limit for the spectral integration is zero for the YY entropy and for the overlap contribution; this is due the distinctive feature of integrable quenches that the corresponding initial states in the post quench basis consists solely of pairs of excitation with opposite momentum. Exploiting translation invariance $\rho$ can be regarded as an even function and we can rewrite the integrals as
\beq
-\frac{L}{2}\int_{-\infty}^{\infty} \text{d}k\, \left\{g(k)\rho(k)- 2\lambda q(k) \rho(k)+ s_{YY}[\rho(k)]\right\}\,,
\eeq
from which the saddle-point equations are obtained via functional differentiation wrt. $\rho$ yielding
\beq
0=g(k)-2\lambda q(k)+\frac{\delta s_{YY}[\rho]}{\delta \rho(k)}\,,
\eeq
which can be recast in the conventional TBA form for the saddle-point density $\rho_\text{sp}$ via the pseudo-energy $\epsilon$ as
\beq
\epsilon(k)=g(k)-2\lambda q(k)-\varphi \star \log\left(1+e^{-\epsilon}\right)\,.
\label{QA2BetaEquationTBAForm}
\eeq
The equation above characterizes a GGE via the spectral densities $\epsilon$, from which $\rho_\text{sp}$ can be computed. Indeed, it has been shown that 
\beq
g(k)=-\sum \beta_i q_i(k)
\eeq
for free fermion quenches \cite{Calabrese_2012_CEF2} and for the Lieb-Liniger model and the BEC quench \cite{denardis2014solution,Palmai-Konik}.
That is, the driving term $g(k)-2\lambda q(k)$ must correspond to a GGE with charge $Q$ biased by $2 \lambda$. Since, as already mentioned,  local operators do not shift the saddle-point of the QA functional \cite{Pozsgay_2011_LocalOperators}, the steady-state expectation value of $O$ is determined by $\rho_\text{sp}^{2\lambda}$ where we made it explicit that $\rho_\text{sp}^{2\lambda}$ is associated with the solution of  Eq. \eqref{QA2BetaEquationTBAForm} and accordingly, 
\beq
\begin{split}
&\lim_{t\rightarrow \infty}\lim_{L\rightarrow\infty}\frac{\langle \Psi|O(0,t)e^{\lambda Q}|\Psi\rangle}{\langle \Psi|e^{\lambda Q}|\Psi\rangle}=\\
&=\int \dd k\, O[\rho_\text{sp}^{2 \lambda}(k)]=\frac{1}{Z^{2\lambda}}\Tr \left[O(0,0)e^{2\lambda Q}\rho_\text{GGE}\right]\,,
\end{split}
\eeq
where 
$\rho_\text{GGE}$ is the GGE dictated by integrable quench, i.e., the unmodified integrable initial state $|\Psi\rangle$. It is also immediate to see that using the same reasoning for the nominator
\beq
\begin{split}
&\lim_{t\rightarrow \infty}\lim_{L\rightarrow\infty}\frac{\langle \Psi|e^{\lambda/2 Q} O(0,t)e^{\lambda/2 Q}|\Psi\rangle}{\langle \Psi|e^{\lambda Q}|\Psi\rangle}=\int \dd k\, O[\rho_\text{sp}^{ 2 \lambda}(k)]\\
&=\frac{1}{Z^{2\lambda}}\Tr \left[O(0,0)e^{2\lambda Q}\rho_\text{GGE}\right]=\Tr \left[O(0,0)\rho^{2\lambda}_\text{GGE}\right]\,,
\label{BiasedGGEConvergence}
\end{split}
\eeq
holds as well, as expected, (and where we defined $\rho_\text{GGE}^{2\lambda}=e^{2\lambda Q}\rho_\text{GGE}/Z^{2\lambda}$).

Finally, in light of the above derivation, we can also comment on why the fluctuations in the initial state are also given by the biased GGE, $\rho_\text{GGE}e^{2\lambda Q}$.
This non-trivial finding was demonstrated  in \cite{Bertini_2024_tFCS, FCSQGases}, but here we can give an intuitive order-by-order reasoning.
We consider the time evolution of the integrated charge 
\beq
\begin{split}
f_{\Psi}(\lambda)&=\lim_{X\rightarrow \infty}X^{-1}\ln\langle \Psi | e^{\lambda Q|_0^X(\tau)}|\Psi\rangle+O(\tau/X)\\
&=:f_{\Psi}(\lambda, \tau)+O(\tau/X)\,,
\end{split}
\eeq
where the vanishing correction $O(\tau/X)$ originates from boundary effects: when $X \gg \tau$ and $\tau$ is small (but not too small as we shortly assume), $Q$ is conserved in the bulk of the subsystem $[0,X]$. Differentiating now $f_{\Psi}(\lambda, \tau)$ wrt. $\lambda$, we find that
\beq
\begin{split}
&\partial_\lambda f_{\Psi}(\lambda, \tau)=\\
&=\lim_{X\rightarrow \infty}X^{-1}\frac{\langle \Psi | X q(X/2,0) e^{\lambda Q|_0^X(\tau)}|\Psi\rangle}{\langle \Psi |e^{\lambda Q|_0^X(\tau)}|\Psi\rangle}+O(\tau/X)\\
&=\lim_{X\rightarrow\infty}\frac{\langle \Psi|e^{\lambda/2\, Q} q(0,\tau)e^{\lambda/2\, Q}|\Psi\rangle}{\langle \Psi|e^{\lambda Q}|\Psi\rangle}\\
&=\text{Tr}\left[q(0,0)\rho_\text{GGE}^{2\lambda}\right]\,,
\end{split}
\eeq
where at the last equality we used \eqref{BiasedGGEConvergence} and the fact that local relaxation is expected to occur at intermediate times, not conflicting with sending $X$ to infinity first. That is, extending the reasoning to higher orders exploiting the order of limits, we have that
\beq
f_{\Psi}(\lambda)=f_{W_\Psi}(0)-f_{W_\Psi}(2\lambda)\,.
\eeq

\section{Clustering of the BEC  state}\label{Clustering}
We can comment on the clustering properties of the BEC state,  $|\Psi\rangle_\text{BEC}$. Indeed this state has very mundane, yet not always trivial correlations. In particular it is easy to check that
\beq
\begin{split}
&\!\!\!\phantom{o}_\text{BEC}\langle \Psi | b^\dagger(x)b(y)|\Psi\rangle^c_\text{BEC}=\\
=&\lim_{L,N\rightarrow\infty}\frac{1}{L}\sum_{k,k'} \langle 0 |\frac{(b_{0,L})^N}{\sqrt{N!}}b_{k,L}^\dagger b_{k',L}\frac{(b_{0,L}^\dagger)^N}{\sqrt{N!}}|0\rangle e^{\ri (k'y-kx)} \\
=&\lim_{L,N\rightarrow\infty}\frac{1}{L} \langle 0 |\frac{(b_{0,L})^N}{\sqrt{N!}}b_{0,L}^\dagger b_{0,L}\frac{(b_{0,L}^\dagger)^N}{\sqrt{N!}}|0\rangle\\
&= \lim_{L,N\rightarrow\infty}\frac{N}{L}=d\,. 
\end{split}
\eeq
using that in finite volume $b(x)=\frac{1}{\sqrt{L}}\sum_ke^{ikx}b_{k,L}$, that is, certain correlations do not decay.
However 2pt connected correlation functions of operators of the form $(b^\dagger(x))^n(b(x))^n$ (including possible derivatives as well) such as the particle density ($n=1$) do satisfy the cluster property. For our cases of interest it is sufficient to focus on this class of operators since they do not change particle number which is also conserved  by the dynamics. For the particular case of the density operator, we can write
\beq
\begin{split}
&\phantom{o}_\text{BEC}\langle \Psi | b^\dagger(x)b(x)b^\dagger(y)b(y)|\Psi\rangle_\text{BEC}=\\
=&\!\lim_{L,N\rightarrow\infty}\!\frac{1}{L^2}\!\!\!\!\sum_{k,k',p,p'} \!\!\!\langle 0 |\frac{(b_{0,L})^N}{\sqrt{N!}}b_{k,L}^\dagger b_{k',L} b_{p,L}^\dagger b_{p',L}\frac{(b_{0,L}^\dagger)^N}{\sqrt{N!}}|0\rangle\times\\
&\qquad\qquad\qquad\qquad\qquad\qquad\qquad \times e^{\ri (k'-k)x}e^{\ri (p'-p)y}\\
=&\!\!\lim_{L,N\rightarrow\infty}\!\frac{1}{L^2}\!\sum_{p} \langle 0 |\frac{(b_{0,L})^N}{\sqrt{N!}}b_{0,L}^\dagger b_{p,L} b_{p,L}^\dagger b_{0,L}\frac{(b_{0,L}^\dagger)^N}{\sqrt{N!}}|0\rangle e^{\ri p (x-y)}
\end{split}
\eeq
which further equals
\beq
\begin{split}
&\phantom{o}_\text{BEC}\langle \Psi | b^\dagger(x)b(x)b^\dagger(y)b(y)|\Psi\rangle_\text{BEC}=\\
&=d^2\!+\!\!\!\!\lim_{L,N\rightarrow\infty}\!\frac{1}{L^2}\!\sum_{p\neq0} \langle 0 |\frac{(b_{0,L})^N}{\sqrt{N!}}b_{0,L}^\dagger  b_{0,L}\frac{(b_{0,L}^\dagger)^N}{\sqrt{N!}}|0\rangle e^{\ri p (x-y)}\\
&=d^2\!+\!\!\!\!\lim_{L,N\rightarrow\infty}\!\frac{N}{L^2}\!\sum_{p} e^{\ri p (x-y)}-\frac{N}{L^2}=d^2+d\delta(x-y)\,,
\end{split}
\eeq

that is,
\beq
\phantom{o}_\text{BEC}\langle \Psi | q(x)q(y)|\Psi\rangle^c_\text{BEC}=d \delta(x-y).
\eeq
For correlation functions of $(b^\dagger(x))^n(b(x))^n$ operators analogous calculations predict $d^n \delta^n(x-y)$ behaviors. The higher powers of the Dirac-$\delta$ function can be regularized by point splitting, which results in $\delta$-correlated connected 2pt functions.

\section{The numerically investigated parameter space for curved trajectories}
\label{App:NumericallyInvestigatedCurvedTraj}
In the following table we show at what $c$, $d$ and $\lambda$ values we have repeated the numerical checks for the curved trajectories. We recall that $c$ denotes the interaction strength in the LL model \eqref{eq:Hamiltonian}, and $d$  the density of bosons after the quench \eqref{eq:Psi0LL}, and $\lambda$ is the counting field for the FCS.
We note that for $c=5$, $d=1$ and $\lambda=1.55$, $c=1$, $d=5$ and $\lambda=1.25$ and $c=1$, $d=0.4$ and $\lambda=1.5$ the variations of $1^\text{dr}(\kappa,\xi,\lambda)$ for fixed $\kappa \in [-\Lambda,\Lambda]$ wrt. tuning $xi$ form $-\infty$ to $0$ or from $0$ to $\infty$ are slightly larger than in the other cases (few \% for $\lambda\propto \mathcal{O}(10^{-1})$ and 20-25\% for $\lambda\propto \mathcal{O}(1)$), namely 35-45\%. Additionally for $c=1$, $d=5$, $1^\text{dr}(\kappa\approx0,\xi,\lambda)\approx5-7$. For $c=1$, $d=2$ and $\lambda=2.25$ this variation is even larger and is up to 50\% wrt. the medium value $6$ at $\xi=0$.
These findings mean that at large values of the counting field the approximations \eqref{1drApprox}, \eqref{q-theta-approx} and \eqref{q-theta-straightline-approx} are less accurate, i.e., taking $1^\text{dr}(\kappa\approx0,\xi,\lambda)\approx \text{const}$ and evolving the charge densities along the Euler characteristics, equated to straight trajectories, respectively.

The other findings of the analysis carried out previously, such as the approximately linear dependence of $\Lambda$ on $\lambda$ even for larger counting fields, the unique zero of $\v^\text{eff}$ as a function of $\xi$, as well as the fulfillment of the condition that $\forall \kappa \in[0,\Lambda],\quad |\max_\xi v^\text{eff}(\kappa, \xi)|,  |\min_\xi v^\text{eff}(\kappa, \xi)|<|\xi^*_1| \quad \text{and}\quad |\max_\xi v^\text{eff}(\kappa, \xi)|,  |\min_\xi v^\text{eff}(\kappa, \xi)|<\xi^*_2$ remain valid in the investigated parameter regime. The same statement is true for the findings of the 1st order expansion of the GHD equations, namely $|\xi^*_1|,|\xi^*_2|$ obtained from the appropriate expansion are $\mathcal{O}(1)$ quantities.

\begin{widetext}
\begin{table}[h]
\begin{tabular}{|c|| c| c|c|c|c|c|c|c|c|c|c|c|c|c|c|c|c|c|c|c|c|c|} 
 \hline
 c & 5.0 & 5.0 & 5.0& 5.0& 5.0& 5.0& 5.0 &1.0&1.0&1.0&1.0&1.0&1.0&1.0&1.0&1.0&0.5&0.5&0.5&0.5&0.5&0.5 \\ 
 \hline
 d & 7.5 & 7.5 & 4.0 & 4.0 & 1.0& 1.0& 1.0 & 5.0& 5.0& 5.0& 2.0& 2.0& 2.0& 0.4& 0.4& 0.4&2.1&2.1&2.1&0.7 &0.7 &0.7\\
 \hline
 $\lambda$ & 0.6 & 0.06 & 0.6 &0.06&1.55&0.6&0.06&1.25&0.75&0.075&2.25&0.75&0.075&1.75&0.45&0.045&1.1&0.5&0.1&1.1&0.5&0.1 \\
 \hline

\end{tabular}
\caption{The numerically investigated parameters}
\label{table1}
\end{table}
\end{widetext}

\section{First order expansion of the GHD equations}
\label{FirstOrderGHD}
Whereas it seems plausible that the aforementioned conclusions also hold at any finite but very small $\lambda$-s, repeating the previous numerical study for very small $\lambda$-s is highly inaccurate. However, it must be ensured that curved hydrodynamics trajectories are not only present above a small but finite value of the counting field $\lambda^*$, which could imply the onset of dynamical phase transitions in the FCS. To show that this is not the case and exclude numerical inaccuracies we therefore differentiate the GHD equations wrt. $\lambda$ and study the first order behavior of the effective velocity and its implication on the presence (or absence) of curved trajectories.

Differentiating the dressing equations  \eqref{eq:Dressing} with the the filling function corresponding to the partitioning protocol \eqref{GHDBasicEq1Theta0iSolution}, we yield
\beq
\partial_\lambda f^\text{dr}(\kappa)|_{\lambda=0}=\left(\varphi\star \tilde{\theta}\, f^\text{dr}\right)(\kappa)+\left(\varphi\star \theta\,\partial_\lambda f^\text{dr}|_{\lambda=0}\right)(\kappa)\,
\eeq
where $\theta(\kappa)$ is the filling function of the GGE characterizing the homogeneous quench problem and 
\beq
\tilde{\theta}(\kappa)=\Theta_{H}(\tau-\kappa)\partial_\lambda\theta^{(2\lambda)}(\kappa)|_{\lambda=0}\,,
\eeq
where $\theta^{(2\lambda)}$ is the filling function of  $\rho_\text{GGE}e^{2\lambda Q}$.
Using \eqref{vEff}, we may write the effective velocity as
\beq
v^\text{eff}(\kappa, \tau,\lambda)\approx v_0^\text{eff}(\kappa)+\lambda v_1^\text{eff}(\kappa, \tau)
\eeq
where $v_0^\text{eff}(\kappa)$ denotes the effective velocity of the homogeneous GGE and for convenience we denote its inverse function as $v_{0,\text{inv}}^\text{eff}(\kappa)$, that is, $v_{0,\text{inv}}^\text{eff}(v_0^\text{eff}(\kappa))=\kappa$. Our first task is to link the parameter $\tau$ to the more physical ray variable via $\xi=v^\text{eff}(\tau, \tau,\lambda)$ from which we have that up to first order in $\lambda$
\beq
\begin{split}
\xi(\tau)&=\tau+\lambda v_{0,\text{inv}}^\text{eff}(v_1^\text{eff}(\tau, \tau))\,,\\
\tau(\xi)&=v_{0,\text{inv}}^\text{eff}(\xi)-\lambda \frac{\dd }{\dd \kappa}v_{0,\text{inv}}^\text{eff}(v_1^\text{eff}(\xi, \xi))\,.
\end{split}
\eeq
To proceed we need to identify the NESS, i.e., $\xi=0$, and find the corresponding $\tau_0$ parameter, which is simply
\beq
\tau_0=-\lambda v_{0,\text{inv}}^\text{eff}(v_1^\text{eff}(0,0))\,.
\eeq
and therefore the effective velocity in the NESS is
\beq
v^\text{eff}(\kappa, \xi=0,\lambda)=v^\text{eff}_0(\kappa)+\lambda v^\text{eff}_1(\kappa,0)+\mathcal{O}(\lambda^2)\,.
\eeq
Expanding $v^\text{eff}_0(\kappa)$ and $ v^\text{eff}_1(\kappa,0)$ up to linear order in $\kappa$ we can determine the zero $\kappa^*$ of the velocity and eventually $\Lambda$ identifying the two quantities. Writing that
\beq
v^\text{eff}_0(\kappa)\approx a_0^1\kappa\,,\quad \text{and}\quad v^\text{eff}_1(\kappa,0)\approx  a_1^0+a_1^1\theta
\eeq
we have that $\Lambda=\lambda a_1^1/a_0^1+\mathcal{O}(\lambda^2)$.
The next steps are analogous to the ones discussed above: we need to show that for $\forall \kappa \in [0,\Lambda]$, $v^\text{eff}_0(\kappa)+\lambda v^\text{eff}_1(\kappa,\xi)$ has two zeros $\xi_1^*,\xi_2^*$ and that $|\min_\xi v^\text{eff}(\kappa, \xi)|,|\max_\xi v^\text{eff}(\kappa, \xi)|<|\xi^*_1|,|\xi^*_2|$.
Keeping in mind that $\Lambda \propto\lambda$, we write $\kappa=\lambda \tilde{\kappa}$, where $\tilde{\kappa}\in[0,a_1^1/a_0^1]$ and first identifying the values $\xi^*_1,\xi^*_2$, we use
\beq
v^\text{eff}(\kappa,\xi^*, \lambda)=0 \Rightarrow v^\text{eff}_0(\kappa)+\lambda v^\text{eff}_1(\kappa, v_{0,\text{inv}}^\text{eff}(\xi^*))+\mathcal{O}(\lambda^2)=0\,,
\eeq
from which
\beq
v_{0,\text{inv}}^\text{eff}(\xi^*)=v_{1,\text{inv},\xi}^{\text{eff},\pm}(-v_{0}^\text{eff}(\kappa)/\lambda)\,,
\eeq
where $v_{1,\text{inv},\xi}^{\text{eff},\pm}$ is the inverse function of $v_{1}^\text{eff}$ wrt. the second argument and for positive/negative values therein, respectively. From the equation above it follows that
\beq
\xi^*_{1,2}=v_{0}^\text{eff}\left(v_{1,\text{inv},\xi}^{\text{eff},\pm}(-a_0^1 \tilde{\kappa})\right)\,.
\eeq
Importantly, the values $\xi_1^*$ and $\xi_2^*$ obtained this way are $\mathcal{O}(1)$ numbers and do not depend on $\lambda$. These values must satisfy that
\beq
\begin{split}
&|\xi^*_1|,|\xi^*_2|>|\min_\xi v^\text{eff}(\kappa, \xi)|,|\max_\xi v^\text{eff}(\kappa, \xi)|\\
&|\xi^*_1|,|\xi^*_2|>|\min_\xi v^\text{eff}_0(\lambda \tilde{\kappa})+\lambda v^\text{eff}_1(\lambda \tilde{\kappa}, v_{0,\text{inv}}^\text{eff}(\xi))|,\\
&\qquad\qquad\quad |\max_\xi v^\text{eff}_0(\lambda \tilde{\kappa})+\lambda v^\text{eff}_1(\lambda \tilde{\kappa}, v_{0,\text{inv}}^\text{eff}(\xi))|\\
&\mathcal{O}(1)>\mathcal{O}(\lambda)
\end{split}
\eeq
therefore the curved trajectories are present at any small finite values of the counting field as well. In the particular case of $c=1$ and $d=2$, we have that $\Lambda=0.5102 \lambda$, and that $\xi_1^*=-3.045$ and $\xi_2^*=3.0094$ for $\tilde{\kappa}=0.2\times 0.5102$;  $\xi_1^*=-2.6699$ and $\xi_2^*=2.6263$ for $\tilde{\kappa}=0.4\times 0.5102$;
$\xi_1^*=-2.4068$ and $\xi_2^*=2.3548$ for $\tilde{\kappa}=0.6\times 0.5102$ and
$\xi_1^*=-2.1352$ and $\xi_2^*=2.0724$ for $\tilde{\kappa}=0.8\times 0.5102$.


\end{document}